\documentclass[graybox]{svmult}

\usepackage{type1cm}        
\usepackage{makeidx}         
\usepackage{graphicx}        
\usepackage{multicol}        
\usepackage[bottom]{footmisc}

\usepackage{comment}

\usepackage{newtxtext}       %
\usepackage[varvw]{newtxmath}       

\makeindex             

\begin{document}

\title*{Unveiling the intrinsic dynamics of biological and artificial neural networks: \\ from criticality to optimal representations}
\titlerunning{Intrinsic dynamics of biological and artificial neural networks} 
\author{Guillermo B. Morales,
Serena di Santo, and Miguel A. Mu\~noz}
\institute{ 
Guillermo B. Morales \at 
Departamento de Electromagnetismo y F{\'\i}sica de la Materia and Instituto Carlos I de F{\'\i}sica
  Te\'orica y Computacional. Universidad de Granada.  E-18071,
  Granada, Spain, \\
 \email{guillermobm@onsager.ugr.es}
\and Serena di Santo \at  Departamento de Electromagnetismo y F{\'\i}sica de la Materia and Instituto Carlos I de F{\'\i}sica
  Te\'orica y Computacional. Universidad de Granada.  E-18071,
  Granada, Spain,\\
  \email{serena@onsager.ugr.es} 
  \and{Miguel A. Mu\~noz} \at Departamento de Electromagnetismo y F{\'\i}sica de la Materia and Instituto Carlos I de F{\'\i}sica
  Te\'orica y Computacional. Universidad de Granada.  E-18071,
  Granada, Spain, \\ \email{mamunoz@onsager.ugr.es}
  }
%
%
\maketitle


\abstract{Deciphering the underpinnings of the dynamical processes leading to information transmission, processing, and storing in the brain is a crucial challenge in neuroscience. An inspiring but speculative theoretical idea is that such dynamics should operate at the brink of a phase transition, i.e., at the edge between different collective phases, to entail a rich dynamical repertoire and optimize functional capabilities. In recent years, research guided by the advent of high-throughput data and new theoretical developments has contributed to making a \emph{quantitative} validation of such a hypothesis. Here we review recent advances in this field, stressing our contributions.
In particular, we use data from thousands of individually recorded neurons in the mouse brain and tools such as a phenomenological renormalization group analysis, theory of disordered systems, and random matrix theory. These combined approaches provide novel evidence of quasi-universal scaling and near-critical behavior emerging in different brain regions. Moreover, we design artificial neural networks under the reservoir-computing paradigm and show that their internal dynamical states become near critical when we tune the networks for optimal performance. These results not only open new perspectives for understanding the ultimate principles guiding brain function but also towards the development of brain-inspired, neuromorphic computation.}

\section{Introduction: the critical hypothesis and machine learning at the edge of chaos}
\label{sec:0}
When neuroscientists poke into the electrophysiological processes of the brain — from microscopic individual neurons to large-scale whole-brain measurements— they systematically detect, even under conditions of quiet rest, an ongoing background of noisy and variable neural activity 
\cite{Softky,Arieli,Raichle,Friston,Deco-Jirsa}. The fact that this ceaseless activity is energetically so costly, had researchers long wondering what crucial functionalities it may entail.

Diverse theoretical explanations for such a background state of reverberant activity have been proposed. A prominent one is that of the \emph{balanced state or asynchronous state}, in which excitatory and inhibitory inputs to any given neuron typically cancel with each other, so that the mean input is below the activation threshold; however, local unbalances create a self-sustained fluctuation-driven noisy state, which may entail important functional advantages for information transmission, processing, etc. \cite{Brunel,van,Renart,Jensen,Corral}.

An alternative hypothesis consider that synaptic weights are tuned homeostatically, so that neuronal networks operate close to a \emph{critical point}, i. e. at the edge between two different types of collective behavior \cite{Schuster,Chialvo2010,Taglia,Haimovici,Breakspear-review,Viola-review,Hidalgo,Turrigiano,Ponce,Martinello,Zhou-jointly,Byrne,RMP}. Knowledge from Statistical Physics teaches us that near critical points there are long-range correlations, scale-invariant spatio-temporal patterns, enhanced collective responses,  very rich dynamical repertoires, etc. \cite{Binney,RMP}. Therefore, it was conjectured that the brain, by being near criticality, could extract crucial advantages from such a plethora of spontaneously-generated collective properties \cite{Plenz-functional,Chialvo2010}. To be a bit more specific and make contact with the previous scenario, one can consider a neural network composed of excitatory and inhibitory units; its overall state can be shifted  (by varying, e.g., the intensity of the synaptic strengths) from the previously discussed asynchronous phase to a synchronous one, where collective oscillations emerge \cite{Zhou-Hopf}. Thus, in particular, it has been argued that the brain could operate at the edge of a synchronization phase transition, so that it can jointly exploit the advantages of both asynchronous noisy states (e.g., for information processing) and synchronous ones (e.g., for coherent information transmission through oscillations) \cite{Plenz-synchro1,Villegas,Poil,Zhou-Hopf,Li-Shew}. In more generic terms, criticality emerges whenever the baseline phase loses its stability and shifts continuously to another
phase; thus, it is also referred to as the \emph{edge of instability} \cite{Magnasco1,Magnasco2,morales_quasiuniversal_2023,morales_optimal_2021}.

In parallel, advances  in the area of artificial neural networks (ANNs) and machine learning  have explored, since the seventies, the idea of computation at the  "edge-of chaos" \cite{Langton,Packard,BerNat}:  ANNs can exploit the combined advantages of stability (order) and responsiveness to inputs (disorder) 
when they operate at the borderline between order and "chaos", a regime that has been proven to be optimal from many different information-theoretic and computational perspectives \cite{BerNat,Boedecker}. 
For example, it has been shown that among all \emph{cellular automata} ---i.e. discrete models of computation, composed by networks of binary units with local updating rules--- only the ones posed at the edge of chaos, are capable of implementing computational algorithms \cite{langton}. Moreover, these  ideas have guided the development of powerful and versatile machine-learning algorithms under the paradigm of "reservoir computing" \cite{Maass,Jaeger0,Jaeger,Boedecker,Stieg}.

In this chapter we will present a combined approach to the two previously described frameworks ---biological and artificial neural networks--- putting them under the common lens of the critical hypothesis.  

The first and second section relies on high-throughput spiking data from the activity of thousands of simultaneously-recorded neurons over 16 different regions the mouse brain \cite{Steinmetz2021}. A phenomenological renormalization-group approach then allows to uncover strong evidence of (quasi-universal) scale-invariance across brain regions. Using diverse techniques based on random-matrix theory and theory of disordered systems, one can further ascribe such scale invariance to a close-to-critical dynamics, quantifying for each specific brain region its distance to the edge of instability. 

In the third section we follow the seminal work of Stringer \emph{et al.}. \cite{Stringer2019} showing how recorded patterns of activity in the mouse visual cortex generate internal representations of external stimuli in an optimal way. To find whether such optimal representations are linked to an underlying critical dynamics, we construct Echo State Networks \cite{jaeger__2001,lukosevicius_practical_2012} and trained them to classify images. Surprisingly, only when the network dynamics is tuned close to the edge of instability we recover the same encoding properties observed in the mouse visual cortex, concomitant with maximal classification performance.

All together the results reviewed here provide us with a general standpoint to analyze neural networks under the lenses of criticality and computation at the edge of chaos. Our analyses reveal that close-to-critical behavior, which is found to be ubiquitous across the brain, can also allow for the construction of optimal input representations in both, biological and artificial neural networks.

\section{In search of scale invariance}
\label{sec:1}

In this section, we show how statistical properties of spontaneous neural activity exhibit scale invariance as well as striking similarities across different brain regions in the mouse brain, despite their inherent heterogeneity. The following analyses resort on the phenomenological renormalization group approach formerly devised by Meshulam \emph{et al.}. in \cite{Meshulam2019}. For the sake of simplicity, we skip here some  technical aspects; the interested reader can find all the details in \cite{morales_quasiuniversal_2023}.

\subsection{A phenomenological Renormalization Group to look for scale invariance in real data}
 \label{sec:RG}

Testing whether neural networks operate near a phase transition is a daunting task, not devoid of controversy within the scientific community \cite{Beggs,Destexhe}. As mentioned in the Introduction, empirical support for the criticality hypothesis has typically relied on the observation of scale-free bursts of activity, called “neuronal avalanches” \cite{BP,vivo}, but other approaches exist, including the study of long-range spatio-temporal correlations \cite{palva2013neuronal,poil2012critical}; the search for statistical critical-like patterns of activity \cite{tkavcik2015thermodynamics}; or the analysis of whole-brain models fitted to match empirically observed correlations \cite{cabral2017functional} (see e.g. \cite{Plenz-review,Breakspear-review,Lucilla,Viola-review,Byrne,RMP} for recent reviews).

A concept that is intimately linked with that of criticality is the existence of scale-invariance. At criticality, a system exhibits scale-invariance, meaning that its properties remain unchanged under rescaling, so that the behavior is characterized by universal features that are independent of the specific details or size of the system \cite{RMP}. This is often captured by power-law distributions ---i.e. the mathematical fingerprint of scale invariance \cite{Sornette}--- of diverse quantities.

The presence of scale-invariance or \emph{scaling} at critical points can be understood because of the divergence of the system's correlation length at these points, implying that ---as the critical point is approached--- the system becomes correlated over larger and larger scales.
Notice that the existence of long-range spatio-temporal correlations gives rise to cooperative effects, where fluctuations at one scale influence fluctuations at other scales; in other words there is no privileged scale and the system is thus scale invariant. 

To analyze this type of problems with many scales, the Renormalization Group (RG) emerges as a mathematical tool, widely used in the realm of Statistical Physics, that allows one to study the behavior of a system as we zoom in or out, capturing the universal features that persist across different scales \cite{Wilson,Kadanoff,Binney}. By exploring the properties of the underlying microscopic constituents, the RG enables us to uncover the emergence of scale invariance.  

Typically, an RG analysis involves the construction of effective descriptions of a microscopic system at progressively larger spatial scales. In the original real-space RG, as proposed by Kadanoff for spin-systems, one has to "coarse-grain" neighboring spins into blocks, which are  then clustered into new blocks at successive steps of the RG transformation \cite{Kadanoff}. The problem, when one tries to adapt this methodology to neural data, is that one typically does not know which neurons are "neighbors" of which, as one usually  lacks information about the detailed connectivity patterns at such microscopic levels. Meshulam \emph{et al.}. found a workaround to this problem in what they termed a "phenomenological Renormalization Group" (PRG) approach; their approach led them to find strong evidence of scale-invariant properties in the recorded neural activity of the CA1 region of the mouse hippocampus \cite{Meshulam2019}. 

\begin{figure*}[h]
\centering
\includegraphics[width=\textwidth]{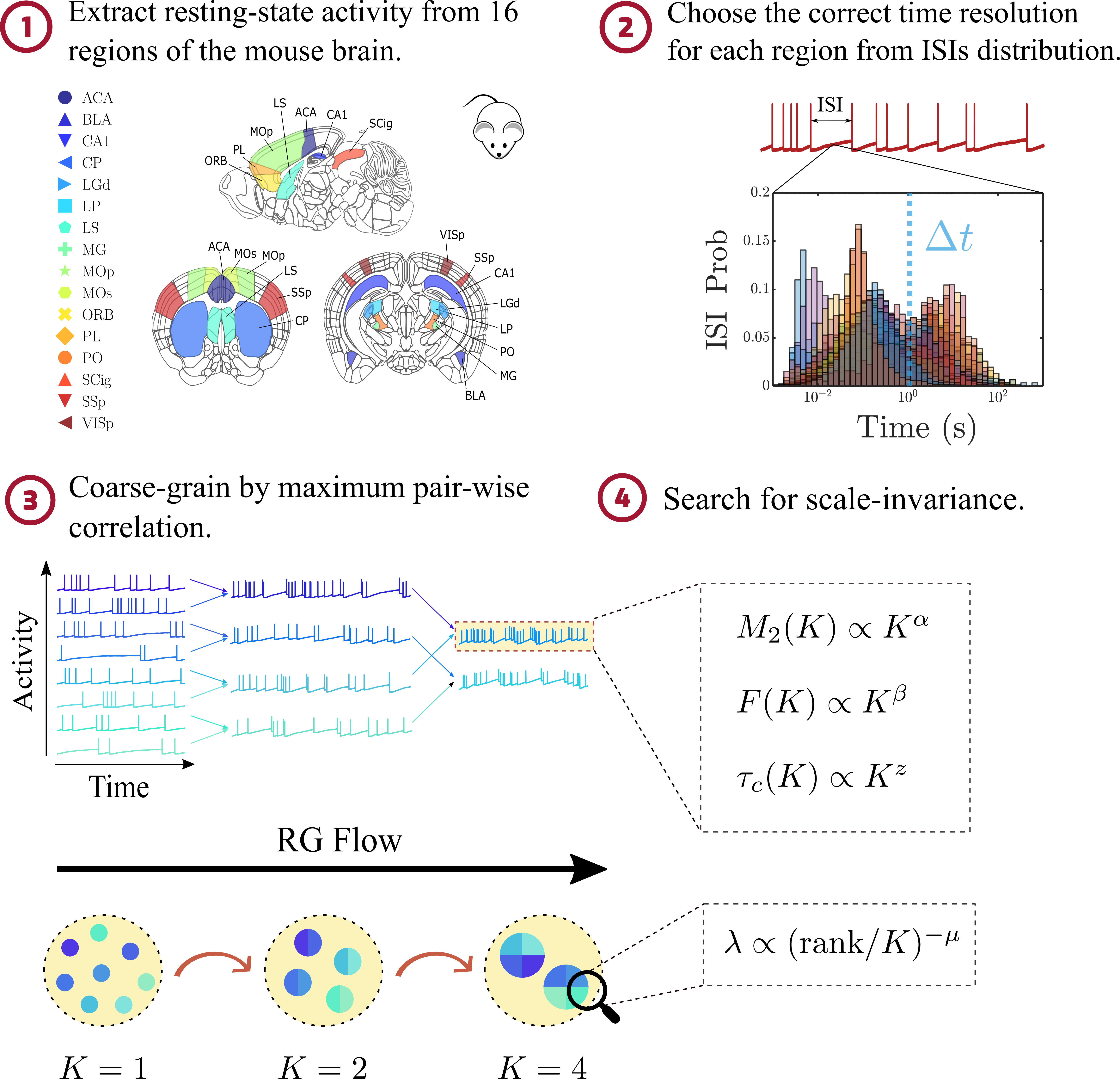}
\caption{\textbf{Workflow of the phenomenological renormalization group analysis}. (\textbf{1}) The raw data is filtered for resting-state activity in regions with at least $N=128$ recorded neurons. (\textbf{2}) For each region, activity is discretized in bins of width $\Delta t$ equal to the geometric mean of the inter-spike-interval (ISI) distribution. (\textbf{3}) The original system is progressively coarse-grained 
by clustering together pairs of most-correlated neurons, so that at step $k$ each new variable contains the summed activity of $K=2^k$ original neurons. (\textbf{4}) By changing the scale $K$, different properties are analyzed as we move towards more coarse-grained descriptions of the system.}
\label{PRG}
\end{figure*}

To understand the context and results of this study, let us consider a set of $N$ neurons, such that the empirically-measured  activity of the $i$-th neuron at a specific time $t_j$ is represented as $x_i(t_j)$, with $j\in[1, T]$, describing a discrete set of non-overlapping time bins up to a maximum time $T$. The coarse-graining procedure then proceeds as follows: instead of grouping neighboring neurons using a criterion of spatial vicinity, one clusters them  based on a criterion of maximal pairwise Pearson's correlation:
\begin{equation}
C_{ij}= 
\frac{\langle \delta x_{i}\delta x_{j}\rangle}{ 
\sqrt{
     \langle       (\delta x_{i} )^{2} \rangle 
\langle (\delta x_{j})^{2}\rangle}},
\end{equation}
where averages are computed across the available discrete time steps and $\delta x$ is the deviation form the mean. More specifically, at the $k$-th coarse-graining step, one selects the two most correlated block-neurons, $i$ and $j_{*i}$, and combine their activities into a new "block-neuron" coarse-grained variable given by:
\begin{equation}
x_i^{(k)} = z_i^{(k)} \left( x_i^{(k-1)} + x_{j_{*i}}^{(k-1)} \right),
\end{equation}
where $k=\{1,...,n_{steps}\}$ indexes the RG step ---one takes $x_i^{(0)}$ to be the original timeseries of the i-th neuron as extracted from the data--- and the normalization factor, $z_i^{(k)}$, is chosen such that the average non-zero activity of the new variables $x_i^{(k)}$ is equal to one. The procedure then is iterated  by considering the second most correlated pair of neurons, and so on, until a set of $N_k=N/2^k$ coarse-grained "block-neurons" is obtained, with each block containing the summed activity of $K = 2^k$ original neurons (see Fig. \ref{PRG} for a schematic representation).

Once the coarse-graining procedure has been completed, the goal is to study how the statistical properties of the system change as one moves progressively across scales. Following \cite{Meshulam2019}, here we focus on four quantities that have a clear interpretation in the classical theory of critical phenomena \cite{Binney}:

\begin{enumerate}

\item The variance of the non-normalized coarse-grained variables. 
\begin{equation}
M_{2}(K)=\dfrac{1}{N_{k}}\sum_{i=1}^{N_{k}}\left[\left\langle \left(\sigma_{i}^{(k)}\right)^{2}\right\rangle -\left\langle \left(\sigma_{i}^{(k)}\right)\right\rangle ^{2}\right]
\end{equation}
where $\sigma_{i}^{(k)}$ is the summed activity of the original variables inside a given cluster and $N_K$ is the number of clusters at scale $k$. Notice that, for totally independent random variables, one would expect the variance to grow linearly in $K$ (i.e., $M_{2}(K)\propto K$), whereas if variables were perfectly correlated $M_{2}(K)\propto K^{2}$. Non-trivial scaling  is therefore characterized by
\begin{equation}
M_{2}(K)\propto K^{\alpha}
\end{equation}
with a certain intermediate value of the exponent $1<\alpha<2$. 

\item The "free-energy" for the coarse-grained variables, defined as:
\begin{equation}
F(K) =-\log\left(S_{K}\right).
\label{Def_FreeEnergy}
\end{equation}
where, $S_{K}$ is the probability that a given coarse-grained neuron is silent at any time step. As  more and more of the initial variables $\sigma_{i}$  are grouped into cluster variables $x_{i}^{(k)}$, one would expect that the probability of having "silent" block-neurons (i.e., the probability that all neurons inside a cluster are silent) decreases rapidly with the size $K$ of the clusters, leading to:
\begin{equation}
F(K)\propto K^{\beta},
\end{equation}
where, in particular, one expects an exponential decay, i.e. $\beta=1$, for initially independent variables \cite{Meshulam2019}.
 
\item The autocorrelation function of the coarse-grained variables. Given that, commonly, fluctuations at larger spatial scales relax with a slower characteristic time scale, one should expect the autocorrelation function of the coarse-grained variables to decay more slowly as one averages over more neurons. In particular, for the $k$-th step of the RG flow, one has:
\begin{equation}
C^{(k)}(t)=\dfrac{1}{N_{k}}\sum_{i=1}^{N_{k}}\frac{\langle  x_{i}^{(k)}(t_{0})x_{i}^{(k)}(t_{0}+t)\rangle - \langle x_{i}^{(k)}\rangle^{2}}
{\langle {(x_{i}^{(k)})}^{2}\rangle - \langle x_{i}^{(k)}\rangle^{2}}
\end{equation}
which, at the steady state, is independent of $t_0$.
Assuming that correlations decay exponentially in time with a characteristic time scale $\tau_{c}^{(k)}$  (i.e., $C^{(k)}(t)=e^{-t/\tau_{c}^{(k)}}$) at each coarse-graining level, dynamical scaling implies that the average correlation function collapses into a single curve when time is re-scaled by the characteristic time scale:
$C^{(k)}(t) = C(t/\tau_{c}^{(k)})$
and that these characteristic times obey scaling with the cluster size:
\begin{equation}
\tau_{c}(K) \propto K^{z},
\end{equation}
where $z$ is the dynamical scaling exponent. 

\item The spectrum of the cluster covariance matrix. Finally, as argued in \cite{Meshulam2019}, if correlations are self-similar across scales, then we should see this by looking inside the clusters of size $K$. In particular, the eigenvalues of the covariance matrix 
must obey a power-law dependence on the fractional rank:
\begin{equation}
\lambda \propto \left(\frac{rank}{K}\right)^{-\mu}.
\end{equation}

\end{enumerate}
Thus, the exponents $\alpha$, $\beta$, $z$, and $\mu$  characterize the possible scale-invariant behavior of the system under study.

\subsection{Quasi-universal scaling exponents across brain regions}
\label{quasi-universal}

By applying the previously introduced PRG to neural recordings (including more than one thousand neurons) of the mouse hippocampus, Meshulam \emph{et al.}. observed the emergence of a non-Gaussian fixed point for the non-zero activity distribution; non-trivial scaling of the variance, "free-energy" and autocorrelation times; as well as spatial scaling reflected in the power-law dependence with the fractional rank for the eigenspectrum of the covariance inside clusters \cite{Meshulam2019}. All this evidence reveals that the underlying neural dynamics is scale invariant (see below for a more detailed explanation).

To analyze the possible robustness or universality of these findings, 
we decided to analyze different brain regions in the mouse brain. In particular, most of the forthcoming analyses rely on the empirical electrophysiological data presented by Steinmetz \emph{et al.}. in \cite{Steinmetz2021}, where the activities of thousands of individuals neurons were simultaneously recorded at a high temporal resolution ($200 Hz$) in several mouse brain regions (see Fig. \ref{PRG}). To conduct a comprehensive analysis, we initially segregated the original data into "resting-state" and "task-related" activity, focusing on the former for the purpose of our study (Fig.1.1).

Before delving any further into the results, let us briefly mention one important aspect that needs to be (and it is rather often not) considered when applying PRG to experimental data: time binning. Typically, studies involving neural recordings of activity, both at the single-neuron level and at lower spatial resolution scales (EEG, fMRI, etc.), involve the discretization of the time series into bins of a certain width $\Delta t$, so that the activity $x_i(t_j)$ of neuron $i$ at bin $j$ is defined as $x_i(t_j)=\int_{t_{j}}^{t_{j}+\Delta t}x_i(t)dt$. Whether the bin size is chosen to match the time resolution of the recording technique, a relevant timescale of the input task, or it is just arbitrarily selected, the truth is that correlations and correlations-based properties of the system can depend drastically on the choice of $\Delta t$. This is especially true for single-neuron activity, given that neurons within the same population can operate at broadly  different time-scales, with spiking frequencies ranging from milliseconds to seconds (see Fig.\ref{PRG}A, and Fig.S4, S5 in \cite{morales_quasiuniversal_2023}). To cope with such heterogeneity, in the forthcoming analyses, the relevant timescale for each brain region (and thus our choice of $\Delta t$) is defined as the \emph{geometric mean} of the inter-spike-interval distribution (see Fig.1.2). As a sanity check, we observed that this choice of $\Delta t$ reproduces to great extent the scaling exponents found in \cite{Meshulam2019} for the mouse hippocampus, using Steinmetz \emph{et al.}'s recordings for this very same area. This result is not trivial and supports the robustness of the observe scale invariance, as the data and recording techniques were very different in both studies.

\begin{figure*}
\centering
\includegraphics[width=10cm]{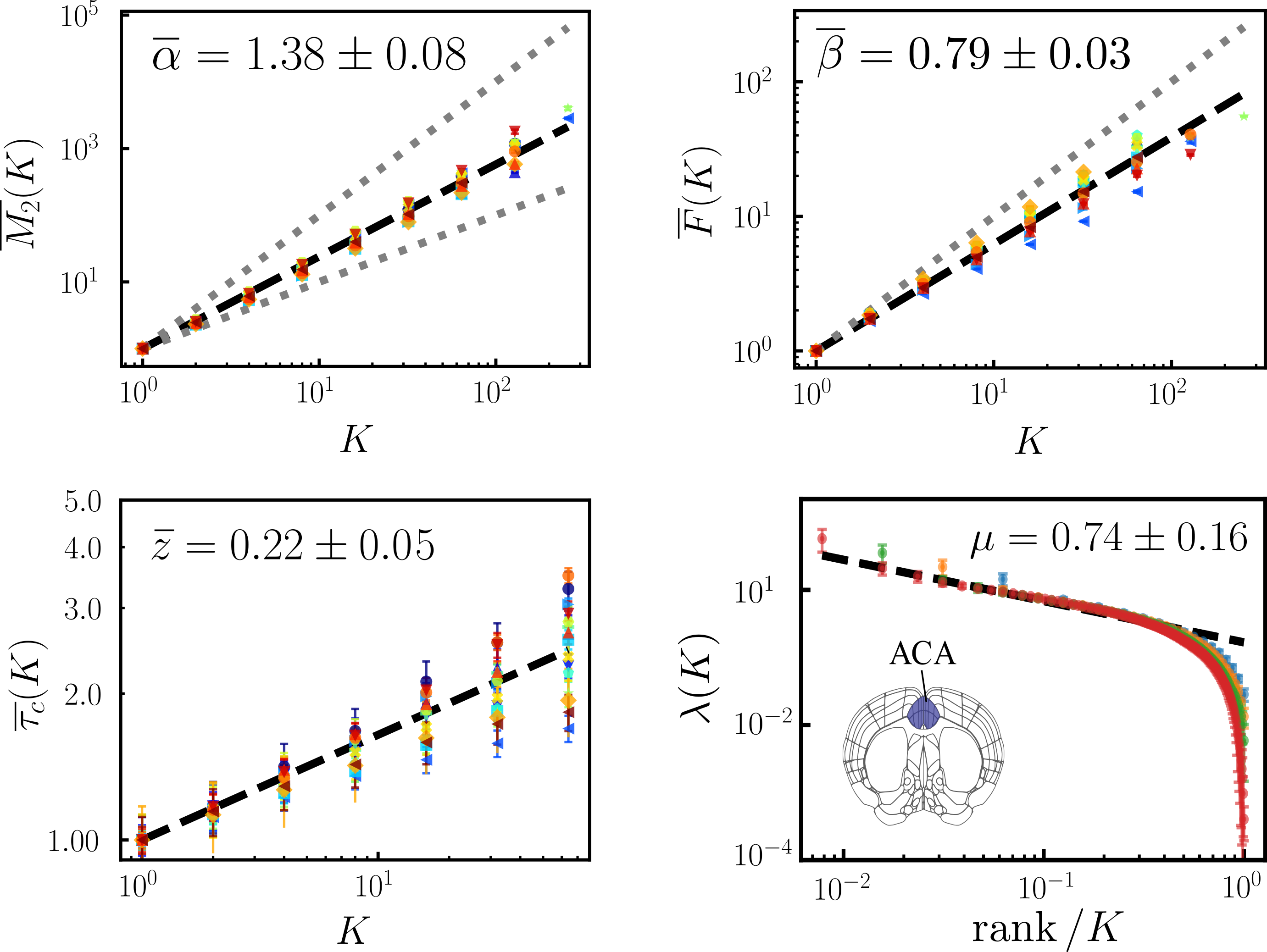}
\caption{\textbf{Phenomenological RG analyses unveils the existence of quasi-universal scaling over 16 different regions of the mouse brain}. Top left panel: Variance of the non-normalized activity as a function of the block-neuron size, $K$, in double logarithmic scale (upper and bottom dotted lines, with slopes $2$ and $1$, mark the fully correlated and independent limit cases, respectively). Top right panel: Scaling of the free energy $F_{k}$,
as defined in Eq.\ref{Def_FreeEnergy} (the dotted line corresponds to the expected behavior for un-correlated variables).
Bottom left panel: Scaling of the characteristic correlation time $\tau_c$ as a function of  $K$ in double logarithmic scale for the different areas. Bottom right panel: Scaling of the covariance matrix spectrum for the resting-state activity in clusters of size ${K=\{16, 32, 64, 128\}}$ (blue, yellow, green and red markers, respectively) for one of the considered brain regions. The rank-ordered eigenvalues decay as a power law of the fractional rank ($\text{rank}/K$) and, even more remarkably, the curves and in particular the cut-offs at different $K$ collapse into the same curve. To facilitate the comparison between areas with different number of neurons, the points of the curves in the first three subplots have been normalized by the value of each quantity at $K=1$ (e.g.,  $\overline{M_2}(K) = M_2(K)/M_2(K=1)$. Errorbars are computed as the standard deviation across split-quarters of data, with lengths typically smaller than the marker size. Adapted with modifications from \cite{morales_quasiuniversal_2023}.}
\label{Fig_2_RG_Scaling}
\end{figure*}  

Fig.\ref{Fig_2_RG_Scaling} summarizes the analysis presented in \cite{morales_quasiuniversal_2023} for the emergence of scale-invariant properties with the RG flow in the recorded activity. In particular, the top left panel, shows an almost perfect scaling of the variance of the coarse-grained variables with an average exponent ${\overline{\alpha}}=1.38 \pm 0.08$ across regions, which is in between the expected one for uncorrelated ($\alpha=1$) and fully-correlated variables ($\alpha=2$). On the other hand, the "free-energy" as defined in Eq.\ref{Def_FreeEnergy}, also exhibits a clear scaling with the  cluster size (Fig.\ref{Fig_2_RG_Scaling}, top right panel), with an average exponent $\overline{\beta} = 0.79 \pm 0.03$ across regions. Finally, temporal and spatial scaling are manifested in the bottom panels of Fig.\ref{Fig_2_RG_Scaling}, respectively, with the former showing the observed dynamical scaling in the decay of the coarse-grained neuron's autocorrelation times $\tau_{c}(K)$, with an average exponent $\overline{z} = 0.22 \pm 0.05$ across regions. For clarity, the power-law decay of the covariance spectrum is plotted only for one of the regions (ACA), where the fractional-rank ($rank/K$) dependence is manifested in the collapse of the curves at different levels of coarse graining. 

To ensure the consistency of the results, we have additionally confirmed that the documented exponent values show minimal variations when the time-discretization bin is changed, with the exception of $\mu$,  which displays an increase during longer time-scales beyond the population activity's typical inter-spike-interval (see Fig. S6 in \cite{morales_quasiuniversal_2023}  for a more in depth discussion).

At the light of the above results, one can confidently state that strong signatures of scale-invariance with quasi-universal exponents are indeed observed across brain regions.

Does such a scale invariance naturally follow from an underlying critical dynamics? The answer to this question is not straightforward as , e.g., an alternative explanation to the observed scale-invariance has been recently derived in terms of latent fields in a model with no critical point in its dynamics \cite{morrell_latent_2021,Nicoletti}. Thus, in the next section we tackle this question not only by looking for signatures of criticality, but actually inferring from the data how far the dynamics lies from a critical point.

\section{Estimating the distance to the edge-of-instability}
\label{sec:edgeofinstability}

Given an empirical system showing some degree of scale invariance, as the previously discussed ones, can we quantify how close it is to the edge of a phase transition or critical point? 

The logic flow we pursue is as follows: 
Let us i)  choose the simplest neuronal model exhibiting a phase transition between two different regimes; ii) introduce two different methods to infer from them the distance to criticality from neural activity measurements; iii) apply and compare the two methods to the recordings in the different regions of the mouse brain analysed above \cite{Steinmetz2021}.

In what follows we employ a simplified linear model. This should not be taken literally as a model for actual coupled neurons but, rather, as an effective model that has been proven to work remarkably well as a linear-response approximation 
of much more complex systems including ingredients such as spiking neurons, non-linear interactions, etc. \cite{Dahmen2019,YuHu2022}.

In particular,  we consider a \emph{linear-rate model} \cite{Crisanti,YuHu2022}, in which the time evolution of $N$ linearly interacting units, each one  described by a continuous variable $x_i(t)$ representing its activity or "firing rate" at instant $t$ and connected with each other through a synaptic-connectivity matrix $J_{ij}$:
\begin{equation}
\dot x_i(t)=-x_i(t) + \sum_{j=1}^N J_{ij} x_j(t)+\xi_i(t).
\label{eq:linearratemodel}
\end{equation}
where $\xi_i(t)$ are  noisy external inputs that,
for simplicity, we assume to be zero-mean uncorrelated Gaussian white noises with $\langle \xi_i(t)\xi_j(t')\rangle= \delta_{ij}\delta(t-t')$.  The synaptic connectivity matrix is assumed to be random, i.e., its elements $J_{ij}$ are sampled from a  Gaussian distribution with zero mean and variance $g^2/N$ (i.e., $J_{ij}\sim \mathcal{N}(0,g^2/N)$). 
\begin{figure}[b]
    \centering
    \includegraphics[width=\textwidth]{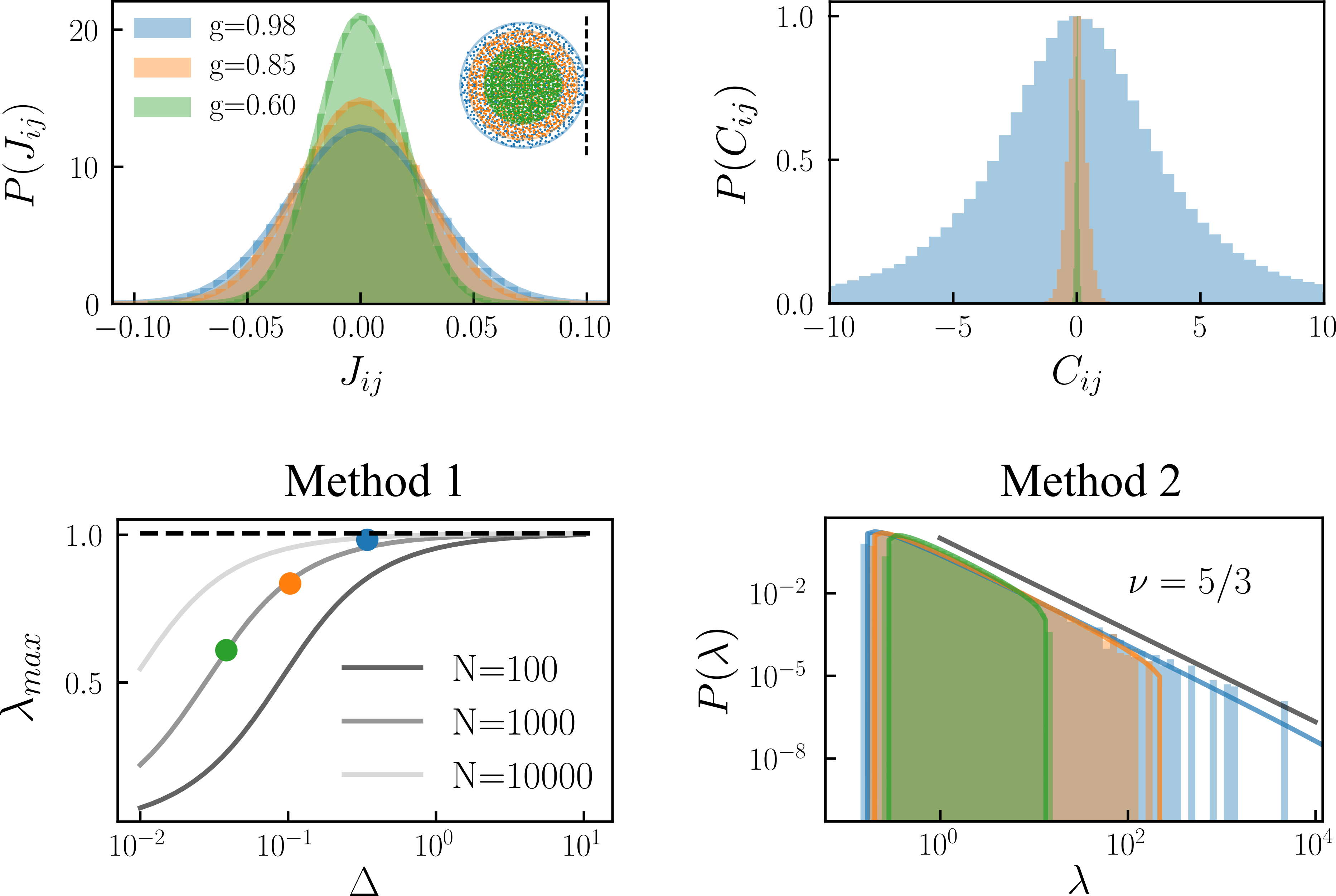}
    \caption{\textbf{Statistics of pairwise covariances provide two different measures of distance to the edge of instability.} For each panel, we compare simulated results for the linear-rate model using 3 different values of $g$. Top left panel: The entries of the connectivity matrix $J_{ij}$ in the linear-rate model follow a Gaussian distribution of zero mean and variance $g^2/N$. The spectrum of the matrix is contained in a circle of radius $g$ in the complex plane (inset), with $\lambda_{max}=1$ marking the edge of instability (black dashed line).
     Top right panel: entries of the sampled spike-count covariance matrix $C_{ij}$ for $N=10^3$. Histograms are re-scaled to their maximum value for visualization purposes. Bottom left panel: maximum eigenvalue of the connectivity matrix $J$ as a function of the relative width of the distribution for the covariance matrix entries. Full lines are drawn using Eq.\ref{eq:lambdaofdelta} for different values of $N$, whereas circles are obtained through simulations of the linear-rate model for the 3 values of $g$ considered (black dashed line represents the edge of instability). Bottom right panel: Distribution of the eigenvalues of the covariance matrix in the linear-rate model. Full lines are plotted using the analytical expression in Eq.\ref{eq:pRC}, whereas histograms are obtained from simulations of the linear-rate model. As $g\rightarrow 1$ the tail of the distribution approaches a power law with exponent $\nu=5/3$ (black continuous line).}
    \label{fig:linearratemodel}
\end{figure}

A well-known result from random-matrix theory, the circular law, asserts that for $N\rightarrow\infty$ a random matrix with Gaussian-distributed entries as above have a spectrum of eigenvalues distributed uniformly over a disc in the complex plane of radius $\lambda_{max}=g$ \cite{Ginibre1965,Mehta1967,Girko2005,Tao2010} (see Fig.\ref{fig:linearratemodel}). Thus, denoting by $I$ the identity matrix, the steady-state of Eq.\ref{eq:linearratemodel} is stable if all the eigenvalues of the matrix $(-I+J)$ are negative, i.e., if $\lambda_{max}=g<1$. In other words, the distance to the edge of instability is fully characterized by the maximum eigenvalue of the connectivity matrix.

Therefore, having access to the full synaptic connectivity matrix in the experimental setup one could automatically have an estimate of the distance to the edge of instability (under the assumption of the underlying linear dynamics). Since this is an extremely daunting task from the experimental point of view \cite{Hage2022,Winding2023}, one needs to resort to alternative strategies, as we will see in the following sections. 

Before closing this section, let us note that the connectivity pattern described by $J_{ij}$ is not realistic from a biological point of view. In particular real neural networks obey Dale's law, meaning that each neuron either excites or inhibits its postsynaptic neurons, which implies that all elements along a column of the connectivity matrix of real neurons should have the same sign. The spectrum of such matrices does not obey the circular law \cite{Rajan}, nevertheless, the results presented below generalize easily to excitatory-inhibitory connectivity matrices \cite{Dahmen2019,YuHu2022}.

\subsection{Empirical correlations and the effective connectivity matrix: a trick from disordered systems}

Typically, one could possibly infer a connectivity matrix from experimental recordings of activity by constructing maximum entropy models that aim to match the observed mean and variance of the empirical correlations (see, e.g., \cite{Cocco} and refs. therein). However, this would imply estimating the $N^2$ entries of the connectivity matrix when in fact, as we saw, the dynamical stability in this simple linear models can be completely determined with just its largest eigenvalue. 
  
Remarkably, an estimation of $\lambda_{max}$ can be obtained from a measure of the pairwise covariance of the activity across neurons. This quantity is much easier to measure experimentally in a robust way, since simultaneous recordings of the activity of large numbers of neurons have become more and more attainable in recent years, thanks to the development of high performance neural probes, as we have already discussed \cite{Steinmetz2021,Demas2021,Zong2022}.

For this, one needs to compute the so-called spike-count covariance matrix (sometimes also called noise covariance or long-time-window covariance). The elements of this matrix are  pairwise covariances of the time-integrated activity across many samples, i.e.: 
\begin{equation}
{C_{ij}=\lim_{\Delta t\rightarrow \infty} \frac{1}{\Delta t}\langle\Delta s_i(t)\Delta s_j (t)\rangle}, \label{long-time_cov}
\end{equation}
where averages are taken over samples and $\Delta s_i(t)$ is computed as:
\begin{equation}
\Delta s_i(t)=\int_t^{t+\Delta t} (x_i(t')-\langle x_i\rangle)dt'.
\end{equation}

Notably, this covariance matrix is linked in a rather model-independent way to the connectivity $J$ through the following relationship \cite{Dahmen2019,YuHu2022,Pernice,Trousdale}:
\begin{equation}
\label{CofJ}
C=(I-J)^{-1}(I-J^T)^{-1}
\end{equation}
with the main assumption that inputs are uncorrelated. 
A simple proof of this equation is presented, for the sake of completeness, in Appendix A.

Note, however,  that one cannot calculate $J$ by inverting Eq.\ref{CofJ} for an empirically measured  $C$ matrix, because the recorded neurons represent only a small subsample of the whole local network so that the inversion strategy is plagued with pitfalls. To circumvent this difficulty,  Dahmen \emph{et al.} recently devised a method to estimate the maximum eigenvalue $\lambda_{max}$ of the connectivity matrix $J$ based on the distribution of the entries of $C$ \cite{Dahmen2019}. More specifically, using ideas and tools from the physics of disordered systems, such as spin glasses, they found the following relationship:
\begin{equation}
\lambda_{max}=\sqrt{1-\sqrt{\frac{1}{1+N\Delta^2}}},
\label{eq:lambdaofdelta}
\end{equation}
where $\Delta=\delta c /\bar{c}$, being $\delta c$ the dispersion of the entries of the spike-count covariance matrix, calculated as the standard deviation of the out-of-diagonal terms of $C$, and $\bar{c}$ the mean variance, i.e. the mean of the diagonal terms. 

Thus, this rather elegant result provides us with an overall measure of the network stability which depends only on the relative dispersion of pairwise covariance values and the system size. Moreover, Dahmen \emph{et al.} 
 show that this result is insensitive to the details of the underlying dynamics and connectivity; for instance they show that Eq.\ref{eq:lambdaofdelta} still holds (as a linear-response approximation) when considering spiking neurons dynamics or excitatory-inhibitory connectivities (see \cite{Dahmen2019} and \cite{Shea-Brown} details).
 
 As we will see in the following section, an alternative estimation of $\lambda_{max}$ can be obtained using the full eigenspectrum of the empirical covariance matrix.

\subsection{Fitting the full covariance matrix spectrum}
\label{yuhu}

In a mathematical tour-de-force, Hu and Sompolinsky took the theory of linear-rate models one step further, deriving an analytical form for the full probability density of the long-time window covariance eigenvalues in the system described by Eq.\ref{eq:linearratemodel} \cite{YuHu2022}. 

\begin{equation}
\label{eq:pRC}
 p(\lambda)=\frac{3^\frac{1}{6}}{2\pi g^2 \lambda^2}\left[\sum_{\xi=1,-1} \xi\left( (1+\frac{g^2}{2})\lambda-\frac{1}{9}+\xi \sqrt{\frac{(1-g^2)^3 \lambda(\lambda_+-\lambda)(\lambda-\lambda_-)}{3}} \right)^\frac{1}{3} \right], 
 \end{equation}
 for $ \lambda_-\leq \lambda \leq \lambda_+,$, with 
 \begin{equation}
 \lambda_{\pm}=\frac{2+5g^2-\frac{g^4}{4}\pm\frac{1}{4}g(8+g^2)^\frac{3}{2}}{2(1-g^2)^3},
 \label{eq:support}
\end{equation}
while $p(\lambda)=0$ for values of $\lambda$ out of the support, i.e. for $\lambda>\lambda_+$ and $\lambda<\lambda_-$.
Observe, in particular, that from Eq.\ref{eq:support} the upper limit of the support $\lambda_+$ diverges
 in the limit $g\rightarrow1$, i.e. close to the edge-of-instability.
 
In fact the distribution $p(\lambda)$ develops a long (power-law) tail of large eigenvalues:
\begin{equation}
    \lim_{g\rightarrow1} p(\lambda) \sim \frac{\sqrt{3}}{2\pi} \lambda^{-\frac{5}{3}}.
\end{equation}
 
Fig.\ref{fig:linearratemodel} shows that the above analytical results, which were derived in the limit of $N\rightarrow \infty$, match accurately the spectrum obtained numerically in simulations of the linear-rate model with $N=10^3$ units.
Thus, in practice, provided one has simultaneously recorded enough neurons (on the order of hundreds) and have enough samples as to meaningfully compute their covariance matrix, it is possible to infer the dynamical regime of the recorded population simply by fitting the sampled covariance matrix eigenvalue spectrum to the theoretical distribution given by Eq.\ref{eq:pRC}, using $g$ as the fitting parameter. This will constitute our second method to estimate the distance to the edge-of-instability from data.

We remark that, although the above expression for the density of eigenvalues was derived assuming a linear-rate model of recurrently connected neurons, Hu and Sompolinsky showed that it provides also an excellent fit to the numerical spectrum of a network of nonlinear rate neurons driven by external noise. Let us caution, in any case, that nonlinearities may introduce a bias in the inferred $g$ value (see \cite{YuHu2022} for more details).

\subsection{Measuring the distance to criticality in mouse brain empirical data}

In this section, we will be applying the two different methods of estimating $\lambda_{max}=g$, as discussed above, to the previously introduced dataset of Steinmetz \emph{et al.} \cite{Steinmetz2021}. To avoid confusion, we will denote by $\hat{g}_d$ and $\hat{g}_s$ the empirical estimates of the distance to criticality obtained using the actual data from the relative dispersion of covariances (first method \cite{Dahmen2019}) and their eigenvalue spectrum (second method \cite{YuHu2022}), respectively. 

\begin{figure}
\centering
\includegraphics[width=\textwidth]{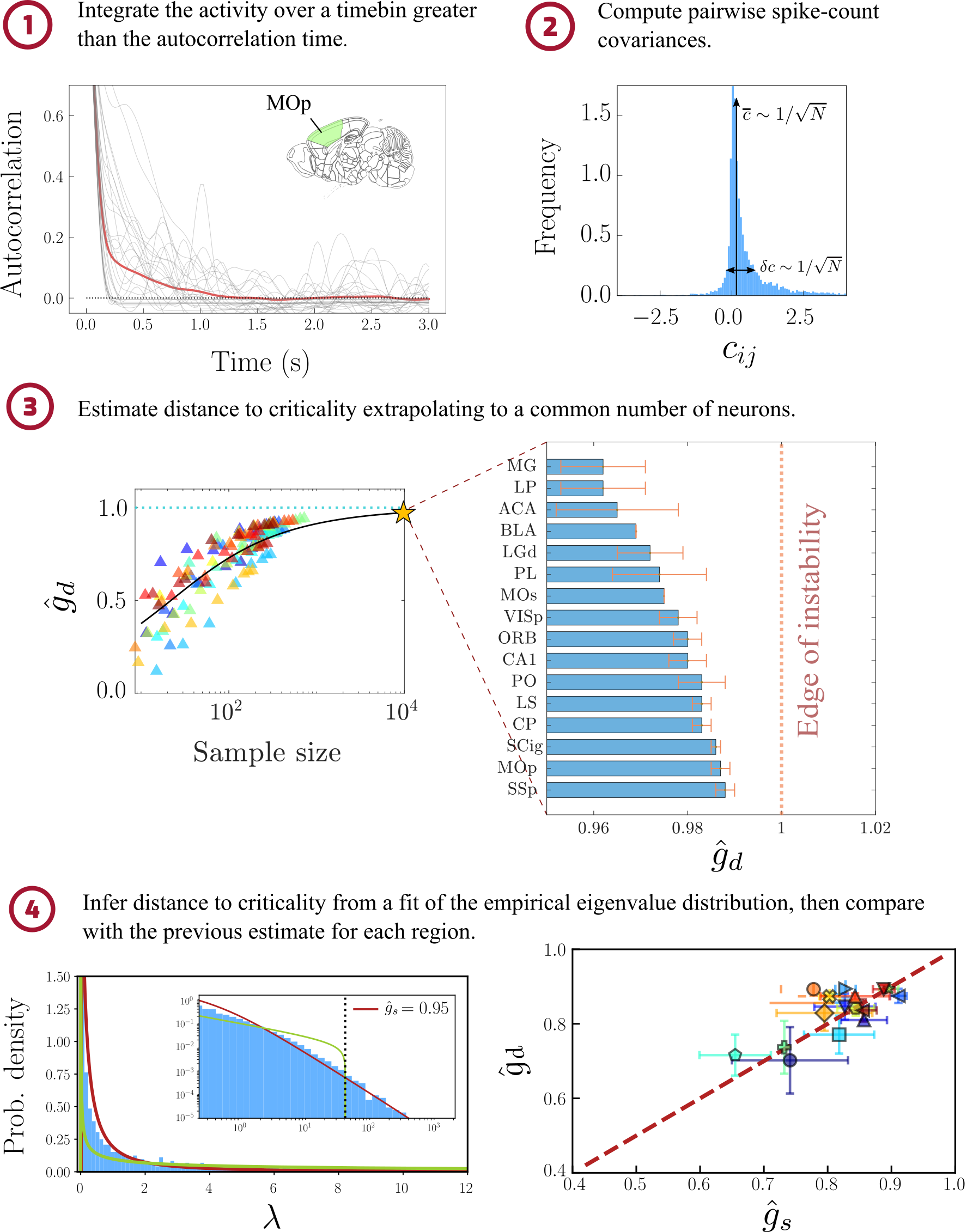}
\caption{\textbf{Distance-to-criticality analysis locates the activity in 16 regions of the brain near the edge-of-instability.}
(\textbf{1}). Decay of autocorrelations in the MOp region (grey lines: $30$ randomly chosen neurons; red line: average over neurons).  (\textbf{2}). Distribution of pairwise covariances in the MOp region, showing a strong peak around $0$. (\textbf{3}). Estimated distance to criticality as a function of the number of neurons $N$ in the sample. For each region (color coded) the points were fitted to Eq.\ref{eq:lambdaofdelta} to obtain an empirical estimate $\hat{g}_d$ of $\lambda_{max}$ at a common number of neurons $N=10^4$. Only an average over such fitted curves (black line) is shown for visualization purposes. (\textbf{4}). Left: Covariance eigenvalues distribution for the MOp region, together with the best-fitting Marchenko-Pastur distribution (green line) and the best-fitting eigenvalue distribution for a linear-rate model of randomly connected neurons (red line, $\hat{g}_s=0.95$). The inset shows a close-up of the corresponding fits in a log-log scale. Right: Estimated values of  $\hat{g}_d$ (dispersion method, \cite{Dahmen2019}) and $\hat{g}_s$ (spectral method, (\cite{YuHu2022}) for each of the 16 regions. In all cases, errors are computed as the SD over different recordings of the same region. Adapted with modifications from \cite{morales_quasiuniversal_2023}.
}

\label{Distance_Criticality}
\end{figure}

First of all, since both methods rely on the computation of the long-time window covariance matrix (Eq.\ref{long-time_cov}), we first need to split the original time series of spiking activity for each neuron into $T$ samples of width $\Delta t$. In practice, the condition $\Delta t \rightarrow \infty$ can be approximated by choosing a window large enough for  autocorrelations to decay (Fig.\ref{Distance_Criticality}.1). In the following results, a characteristic time $\tau$ for each region was obtained from an exponential fit to the mean autocorrelations, then a timebin $\Delta t=1s$ was chosen so that $\Delta t>\tau$ while maximizing the number of samples available. Next we calculate the spike-counts for each neuron by integrating the activity within the previously defined time windows and compute the covariance matrix across neurons (the distribution of entries of such a matrix is shown in Fig.\ref{Distance_Criticality}.2 for one of the regions.). 

A first estimate of the distance to criticality in each region is obtained using Eq.\ref{eq:lambdaofdelta} (see Fig.\ref{Distance_Criticality}.3). By taking random subsamples of the recorded neurons, one can generate a curve of the estimated maximal eigenvalue $\hat{g}_d$ as a function of network size. Then, to make a sensible comparison between regions with different number of recorded neurons,  we extrapolate the value of $\hat{g}_d$ for a common number of neurons $N=10^4$ in each area (Fig.\ref{Distance_Criticality}.3, left). As we can observe in Fig.\ref{Distance_Criticality}.3 (right), all considered areas turn out to be very close to the edge of instability.

Alternatively, one can infer $g$ from the best-fitting parameter $\hat{g}_s$ minimizing the Cramer von-Mises distance between the empirically obtained cumulative distribution and the theoretically-derived one $F(\lambda)=\int_{-\infty}^\lambda p(\lambda)d\lambda$ \cite{YuHu2022}. To illustrate this, in Fig.\ref{Distance_Criticality}.4 we plot the empirical distribution of covariance eigenvalues for an example region, together with the best fitting distribution $p(\lambda)$ (obtained for $\hat{g}_s=0.95$, very close to the critical value). Moreover, we report for comparison the best-fitting Marchenko-Pastur eigenvalue distribution (i.e., the one expected for uncorrelated random variables), which clearly fails to capture the long tail of the empirical eigenvalue distribution.

To summarize, Fig.\ref{Distance_Criticality}.4 presents a comparison of the two alternative estimates of the distance to criticality, $\hat{g}_d$ and $\hat{g}_s$, for all the $16$ considered regions of the mouse brain, revealing that ---within errorbars--- they are typically in excellent agreement. Thus, the above results let us conclude that all the analyzed regions are, to greater or lesser extent, close to the edge of instability. Understanding the origin and meaning of the existent variability across regions is beyond our scope here.

\subsection{Closing the loop: do critical models show scale invariance?}

In previous sections we have described a PRG
scheme can be used to reveal genuine scale invariance in a dataset of simultaneous neuronal recordings (Section \ref{sec:RG}) as well as two different ways to infer how close/how far to the edge of instability these systems are posed, both based on measures of the covariance of the units' activity (Section \ref{sec:edgeofinstability}). The missing 
link is that, thus far, we do not have evidence that systems ---in particular theoretical models--- posed near the edge of instability give rise to scaling behavior compatible with that observed in actual neural data.

 Therefore, in order to complete the picture, here we apply the PRG procedure to the linear-rate model (Eq. \ref{eq:linearratemodel}), for which we can control the distance to the edge of instability through the coupling parameter $g$. We \emph{a priori} expect to find non-trivial scaling for $g\simeq1$ but not for $g<<1$ and, moreover, we wonder whether the scaling exponents are similar to the empirically measured ones.

The results of these analyses are illustrated in Fig.\ref{fig:RGYuhu}; in particular they  reveal that: 

\begin{figure}[b]
    \centering
    \includegraphics[width=0.95\textwidth]{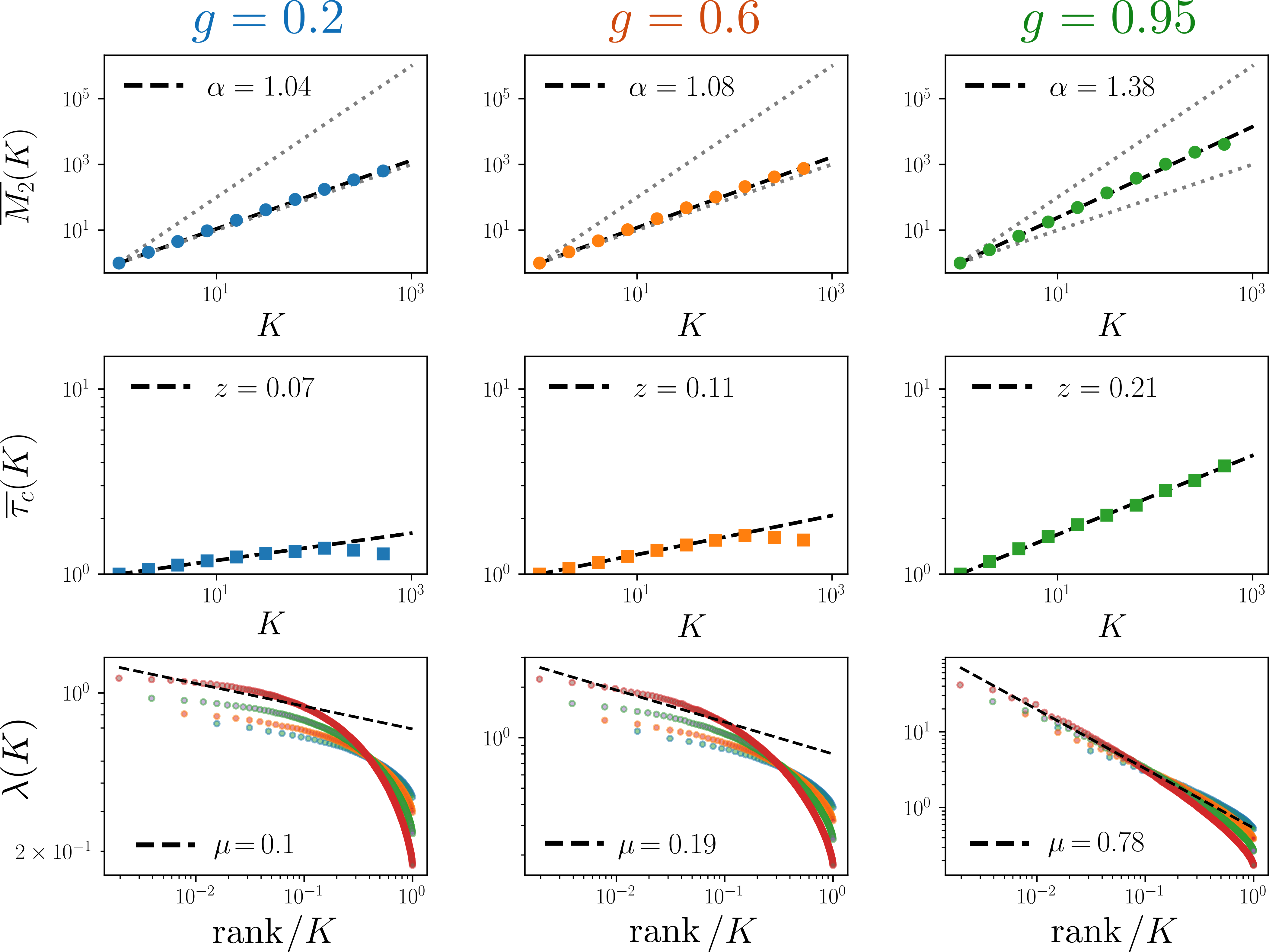}
    \caption{\textbf{A phenomenological RG analysis for the linear-rate model shows signatures of scale-invariance when the system is close to the edge-of-instability.} RG over a randomly recurrent network of linear-rate neurons for different values of the overall coupling strength $g$. Top: Variance of the non-normalized activity, Middle: scaling of the characteristic correlation time, Bottom: Covariance matrix spectrum for the activity in clusters of size $K$, as a function of the block-neuron size. The system closest to the edge-of-instability ($g=0.95$) shows scalings and power-law exponents similar to the ones observed in real data. The
    method also detects that further away from the edge-of-instability the model lacks of non-genuine scale invariance: i) the variance of activity shows a scaling exponent compatible with the one for uncorrelated variables; ii) the rank-ordered spectrum is close-to-flat and shows a cutoff for large cluster sizes, and iii) the collapse across RG steps of the rank-ordered covariance matrix spectrum is lost. Parameter values: $N=1024$, $\sigma=1$, $\tau=1$. Adapted with modifications from \cite{morales_quasiuniversal_2023}.}.
    \label{fig:RGYuhu}
\end{figure}

\begin{itemize}

\item The variance of the activity scales with cluster size with a trivial exponent $\alpha\simeq1$ for $g=0.2, 0.6$, expected for independent units; however, for $g=0.95$, ---sufficiently close to the limit of instability--- it shows the non-trivial exponent $\alpha=1.38$, in surpriseingly good agreement with the average exponent measured for the spiking neurons ($\bar\alpha=1.38\pm0.08$, Fig.\ref{Fig_2_RG_Scaling}, top right panel). 
 
 \item The characteristic time for the autocorrelation function shows a disruption of scaling for $g=0.2,0.6$ for large cluster sizes and a close-to-trivial scaling for small cluster sizes; however, for $g=0.95$ the scaling holds for the all the considered cluster sizes and the measured dynamical scaling exponent is very close to average experimental value ($\bar z=0.22\pm0.05$, Fig.\ref{Fig_2_RG_Scaling}, bottom left panel).

 \item The rank-ordering of the spectrum of eigenvalues of the covariance matrix has a cut-off that changes with cluster-size quite dramatically for $g=0.2,0.6$. On the contrary, for $g=0.95$ there is a ---not perfect but much better--- collapse of the curves at different levels of coarse-graining and a power-law trend with an exponent close to the value measured across actual brain regions ($\bar\mu=0.84\pm0.14$, Fig.\ref{Fig_2_RG_Scaling}, bottom right panel shows the collapse for one example region).

\end{itemize}

Summing up, we find that a very simple linear-rate model tuned to the vicinity of its critical point generates scale-invariant patterns of activity, with
associated critical exponents that match remarkably well those observed for the resting-state activity across brain regions in the mouse brain. Understanding whether more elaborate models, including e.g. the Dale's rule in their connectivity matrix, non-linear interactions, spiking neurons, etc.
do behave in the same universal way or, rather, represent a better description of the scaling features in actual neural networks in the brain remains as an open task for the future.

\section{Representing the external world}

In the previous sections, we have relied on recordings of spontaneous (resting-state)  activity to show, through different arguments, how the dynamics of these networks seem to be poised near a phase transition. However, brains rarely work in such a resting state, but are rather constantly faced with external stimuli that need to be processed. 

One needs to be particularly cautious when trying to apply the techniques presented in the previous sections to recordings of input-evoked activity. For instance, the theory that allowed us to estimate the distance to criticality of different brain regions in Section 3.3 requires the time-series to be stationary, a property that usually does not hold in input-evoked activity. How much can then be said about the dynamical regime of this type of activity?

To answer this and related questions, we first introduce a method that allows one to separate input-related activity ---this is, activity that codifies information about the presented external stimulus--- from the "background" activity, which, for practical purposes, we identify here with noise (although, of course, it could emerge from the encoding of other unknown or latent processes). 

\subsection{Disentangling input-related and background activity}

When trying to make sense of high-dimensional data, as in the case of recordings of thousands of neurons, a common and very practical tool to reduce the dimensionality of the problem and find directions of interest is Principal Component Analysis (PCA). Mathematically speaking, given a matrix of observations $X\in\mathbb{R}^{N\times T}$, where $N$ is the number of variables and $T$ the number of samples or observations, PCA seeks to find a transformation or change of basis, $P$, such that $Y=PX$ is a better representation of the data, meaning that: i) the redundancy between the variables (i.e., the covariance) is minimized, and ii) the signal-to-noise ratio is maximized, this is, $P$ projects the original data into the directions of maximal variance. PCA finds a particularly elegant solution: in the new basis, the covariance matrix $C_{Y}$ must be diagonal, so the projector $P$ is just the matrix of eigenvectors of $C_{X}$ arranged in columns \cite{PCA}. 

Let us now introduce the problem set up by Stringer \emph{et al.} in \cite{Stringer2019}, where they recorded the response of $N\approx10000$ neurons in the mouse visual (V1) cortex to the presentation of $T=2800$ images. If PCA was applied directly over this dataset, one would find the directions of maximal variability, but nothing could be said about the origin of this \emph{overall} variance. In fact, it is estimated that half of the variance of the visual-cortex activity is unrelated to the presented stimuli. Is there a way to tell apart the variance in neural activity steaming from the stimulus encoding ("signal variance"), from the intrinsic or trial-to-trial variability ("noise variance")?

In short, Stringer \emph{et al.}. showed that the stimulus-related variance can indeed be extracted from empirical data by measuring the amount of shared variance between the neural responses to a repeated presentation of the same set of stimuli \cite{Stringer2019}. The method can then be easily extended to extract not only the signal, but also the noise variance, assuming they span orthogonal sub-spaces within the N-dimensional space \cite{morales_quasiuniversal_2023}. We leave the mathematical details of this method, termed cross-validated PCA (cvPCA), to Appendix C.

Since the above cvPCA gives us not only the eigenvalues (input-related variance), but also the associated eigenvectors, it is then possible to project the original activity into an input-related subspace $\Psi$, independent of the trial, and an orthogonal, trial-dependent subspace $\Sigma_{k}$, such that on trial $k$:
\begin{equation}
    X_{k}=\Psi + \Sigma_{k}.
\end{equation}

Equipped with this method allowing us to disentangle input-related variance of the activity from background fluctuations, let us now see how properties of the former can give us information about the geometry of the internal neural representations.

\subsection{A smooth representation of external stimuli}

One way of understanding how the brain encodes information about the external world within the activity of thousands of neurons  is through the idea of "neural representations". Given a set of $N$ neurons that we know are involved in the processing of a certain type of stimulus, their average response to one of such inputs can then be understood as a point in the N-dimensional phase space spanning all the possible states of activity. 

In a similar way, a set of $T$ similar or related stimuli could be mapped into $T$ points spanning a manifold of dimension $D\leq N$, since one would expect that correlations between the response of the neurons to similar inputs would generate some redundancy in the encoding: i.e., their activities are not expected to be randomly distributed in the whole $N$-dimensional space, but to occupy a lower-dimensional "representation" manifold \cite{Chung,Ganguli}.

In particular, for a neural encoding to be robust ---meaning that representations do not change drastically against small variations of the inputs or perturbations--- it is reasonable to assume that these neural manifolds need to be continuous and differentiable (henceforth "smooth"). Remarkably, the authors in \cite{Stringer2019} presented a mathematical proof that conditioned the "smoothness" of such manifolds to the spectrum of the covariance matrix of neuronal responses to a set of stimulus. More specifically: the eigenvalues of the covariance matrix, when ranked from the largest to the smallest, should decay with their rank according to a power law with an exponent 
\begin{equation}
    \mu \geq 1 + 2/d,
\end{equation}
where $d$ represents the embedding dimension of the inputs; if this happens the representation is "smooth", and the other way around (details of this proof are not presented here;  the interested reader is referred to \cite{Stringer2019} for an in-depth explanation of this smoothness-to-spectrum relation and its implications). Thus, it is straightforward to see that $\mu = 1$ serves as a lower bound for the eigenspectrum decay-exponent for complex, high-dimensional inputs.

Experimental recordings of over $10000$ individual neurons in the mouse visual cortex exposed to a vast sequence of images with different dimensionality have validated these theoretical predictions in a remarkable way \cite{Stringer2019}. In particular, Stringer \emph{et. al.} showed that the brain is capable of generating internal representations which are optimal in the sense that they are as high-dimensional as possible (i.e., with a slow-decaying covariance spectrum, meaning that there are many principal components to be used for information encoding), while respecting the aforementioned boundary for the smoothness of the representation: $\mu \sim 1 + 2/d$.

\subsection{Optimal representations at the edge of instability: a reservoir-computing example}
We have argued that the mouse brain is capable of achieving optimal internal representations of external stimuli, provided these live in smooth manifolds which are as high-dimensional as possible given the constraint presented in the previous Section. But, what does this optimality principle says about the dynamics underlying neural representations? Does it have any relationship or dependency on underlying critical behavior?

Before trying to answer the above questions, let us remark that there are two caveats preventing us from using the results and methods discussed in the previous sections (Sec.3.1 and Sec.3.2) for estimating the distance to the edge-of-instability in recordings of input-evoked activity. 

\begin{enumerate}
    \item The aforementioned methods require the timeseries to be stationary, a condition that typically does not hold if the recordings of neural activity are not from resting-state networks. 

\item Thus far our estimates of the distance to criticality have relied on measures of the degree to which fluctuations of the resting-state activity of neurons are shared (known as spike-count or \textbf{noise correlations}), whereas the smoothness condition derived in \cite{Stringer2019} refers to the spectrum of the \textbf{signal correlations}, i.e., the pairwise correlations in the mean responses of neurons to a given set of stimuli. 
\end{enumerate}

For these reasons, we need to take a different approach; namely, we will try to recover the phenomenology observed in \cite{Stringer2019} for neurons in the mouse visual cortex by designing an artificial neural network capable of performing an image-classification task. The interested reader can find detailed explanations of all the forthcoming results in \cite{morales_optimal_2021}.

\begin{figure}
\begin{centering}
\includegraphics[width=\textwidth]{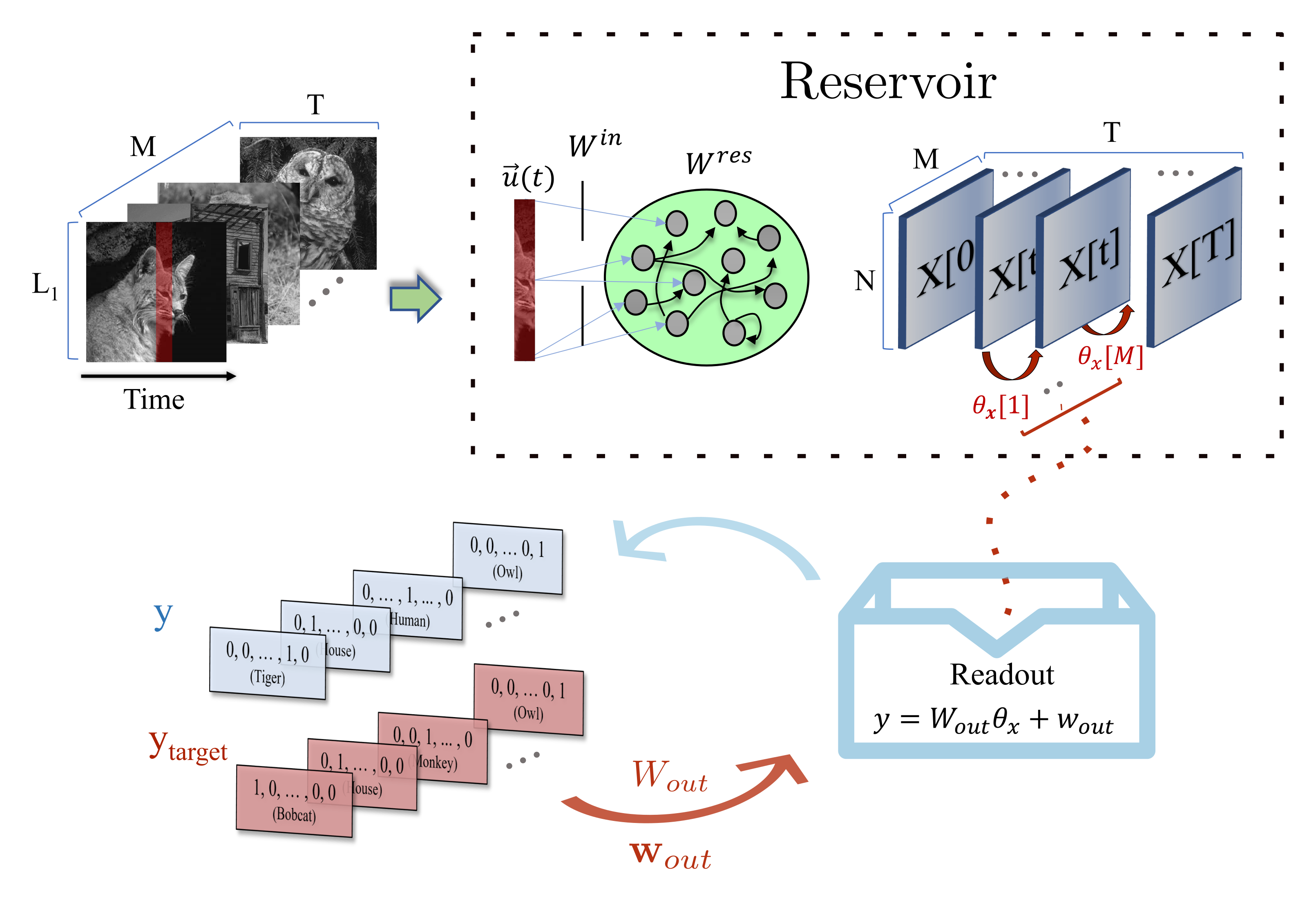}
\par\end{centering}
\caption{{\bf Training process of an ESN for an image classification task.} Images are converted to multivariate time series and then fed into the reservoir. For each processed image a set of parameters $\theta_{x}$ is generated, which characterizes the high-dimensional state of the reservoir, i.e. the ``\emph{reservoir model space}''. Those parameters are then fed into the readout module, that linearly transforms the information in the reservoir model space into an output label. Finally, output weights $W_{out}$ and biases $\mathbf{b}_{out}$ are generated by minimizing the error between the predicted and target labels. Red arrows indicate steps in which a Ridge regression is performed. Adapted with modifications from \cite{morales_optimal_2021}.}\label{Figure_0}
\end{figure}

The following analyses resort on a paradigmatic example of reservoir computing: the Echo State Network (ESN) \cite{jaeger__2001, lukosevicius_practical_2012, reservoir}, whose architecture and training for image classification is summarized in Fig.\ref{Figure_0}. 
More specifically, the basic structure of ESNs consists of: 

\begin{itemize}

\item An input layer, which scales a number $L_{1}$ of inputs at each time step before they arrive in the reservoir, according to some random weights  $W^{in}\in\mathbb{R}^{N\times L_{1}}$.

\item A reservoir consisting of $N$ internal neurons/units, connected with random weights  $W^{res}\in\mathbb{R}^{N\times N}$, whose states evolve according to a non-linear, time-discrete dynamical equation under the influence of a time-dependent input. In this way, the reservoir projects the external input into the high-dimensional space spanned by the activity of its internal units.

\item An output layer, that converts the information contained in the high-dimensional states of the neurons (which serve as an internal representation of the inputs) to generate the final output.
\end{itemize}

Unlike other artificial neural networks, the internal weights or "synaptic connections" in ESNs do not need to be updated during the learning process. Training is achieved by just modifying the layer of output weights that readout the network internal states. As we will see in what follows, the internal dynamics of ESNs can be found either in a stable or in a chaotic regime ---with a critical point in between--- depending on the hyper-parameters of the network (see below). 

In order to adapt the ESN architecture ---usually employed in time-series analyses--- for an image classification task, we used black and white images with $L_{1}\times L_{2}$ pixels (each of them characterized by a value in the $[0,1]$ interval, representing a normalized gray-scale) and  converted them into multivariate time series by considering their vertical dimension as a vector of  $L_{1}$ elements or features, that ``evolve" along $T=L_{2}$ discrete ``time" steps. 

One can then define a standard training protocol \cite{bianchi_reservoir_2021} in which, as illustrated in Fig.\ref{Figure_0}, at each time $t \in[0,T]$, vectors $\mathbf{u}(t) \in[0,1]^{L_{1}}$ corresponding to columns of the image are fed as inputs to the ESN. In this way, the network dynamics for the reservoir states is  given by the following non-linear activation function: 
\begin{equation}
\mathbf{x}(t)=\tanh(\varepsilon W^{in}\mathbf{u}(t)+W^{res}\mathbf{x}(t-1)) + \mathbf{\xi}(t),\label{eq:States_Update}
\end{equation}
where $\varepsilon$ is an input scaling factor and $\mathbf{\xi}(t)$ is a white-noise term with zero mean and correlation $\langle \xi_i(t)\xi_j(t')\rangle=\sigma^2 \delta_{ij}\delta(t-t')$. Unless otherwise stated, we always consider the deterministic case in which $\sigma=0$.

Using a supervised learning scheme, the goal of the ESN is to transform (map) the internal representations of the input into an output label $\mathbf{y}\in\mathbb{\mathbb{N}}^{F}$ that correctly classifies each image in the test set as belonging to one of the $F$ existing categories or classes:
\begin{equation}
\mathbf{y}=W_{out}\theta_{x}+\mathbf{b}_{out},\label{Eq_Readout}
\end{equation}
where $W_{out}\in\mathbb{\mathbb{R}}^{F\times N(N+1)}$ and $\mathbf{b}_{out}\in\mathbb{\mathbb{R}}^{F}$,  defined as the output weights and biases, are determined through a Ridge regression that minimizes the error between the produced and target label for all the presented images in the training set. On the other hand,  $\theta_{x}=[\text{vec}(W_{x}) ; \textbf{b}_{x}]$ is the ``reservoir model space", a set of parameters encoding the reservoir \emph{dynamical} state for a given input (image), obtained from a linear regression to predict the next reservoir state from the past one at discrete time steps:
\begin{equation}
\mathbf{x}(t+1)=W_{x}\mathbf{x}(t)+\mathbf{b}_{x}.
\end{equation}

Summarizing, once trained, our ESN is able to decode the identity of the presented input images from the internal representations elicited by such inputs in the reservoir. Thus, a direct comparison with the experiment in Stringer \emph{et al.} naturally raises the following questions: 

\begin{itemize}
    \item Are ESNs able to generate ``optimal" internal representations of d-dimensional inputs, which are as high-dimensional as possible while respecting the smoothness boundary given by $\mu > 1 + 2/d$?

    \item If so, what characterizes the dynamical regime in which these optimal representations emerge?

    \item  Are the dimensionality and smoothness of the internal representations related to the network's ability to classify the input images?
\end{itemize}

In order to answer the first two questions, we modified two hyperparameters that drastically determine the dynamical regime within the reservoir: the \emph{spectral radius} ---largest eigenvalue--- of the internal connectivity matrix, which controls the dynamical stability inside the reservoir when no input is fed into the network, and the \emph{scaling factor} $\varepsilon$ of the input weights, which can turn an initially expanding mapping into a contracting dynamics, as stronger inputs tend to push the activities of the reservoir units towards the tails of the non-linearity. All the other parameters (number of units, sparsity of connections, weight distribution, etc) were kept at fixed values (see \cite{morales_optimal_2021} for further details). 

Having set the two knobs that tune the dynamical regime of the ESN, one just needs to define a good dial to figure out how close or far the dynamics is from the edge-of-instability. For this, we used the maximal Lyapunov exponent (MLE), a typical measure of the system sensitivity to perturbations (and, by extension, a way of quantifying chaos), defined as:
 \begin{equation}
MLE=\underset{k\rightarrow\infty}{lim}\dfrac{1}{k}log\left(\dfrac{\gamma_{k}}{\gamma_{0}}\right)
\end{equation}
where $\gamma_{0}$ is the initial distance between the perturbed and unperturbed trajectories, and $\gamma_{k}$ is the distance between the trajectories at time-step $k$ \cite{boedecker_information_2011, sprott_chaos_2003}. Thus, a positive value of the MLE is characteristic of chaotic dynamics, whereas the system is said to be stable to local perturbations provided $MLE<0$.

\begin{figure}
\centering{}\includegraphics[width=\textwidth]{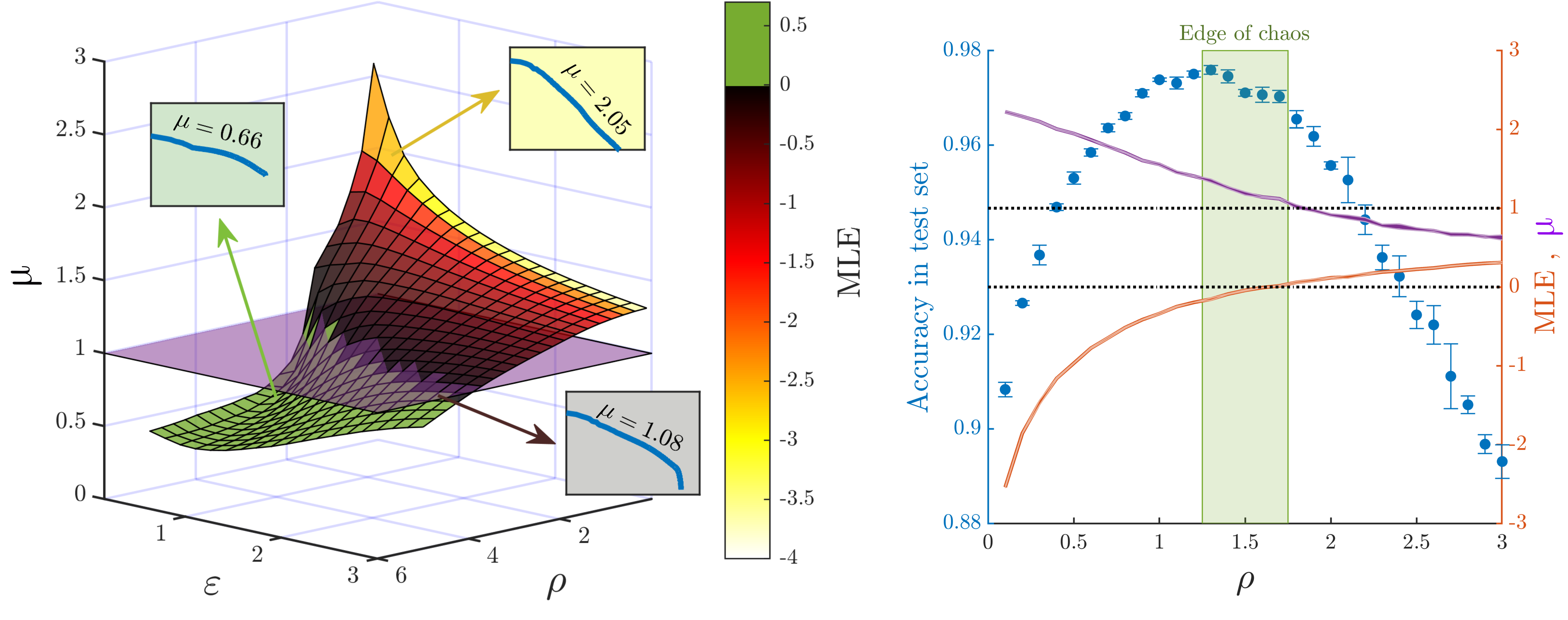}
\caption{\textbf{Optimal performance and smooth, high-dimensional internal representations meet near the edge-of-instability. }Left: exponent for the power-law decay of the spectrum of the activity covariance matrix  as a function of the spectral radius ($\rho$) and input scaling factor ($\varepsilon$) of the reservoir, plotted together with the maximum Lyapunov exponent (MLE) color-coded within the surface. The insets correspond to the activity covariance matrix eigenspectrum measured in three different points of the parameter space, where the variance in the n-th dimension (n-th eigenvalue) scales as a power-law  $n^{-\mu}$  of the rank.  The purple plane marks the boundary $\mu =1$  for smooth representations of high-dimensional inputs. Right: accuracy in MNIST testset (blue dots), maximum Lyapunov exponent (orange line) and best fit exponent for the power-law spectrum of the activity covariance matrix (purple line). Errors were estimated as the standard deviation over ten different realizations of the ESN. Adapted with modifications from \cite{morales_optimal_2021}.}\label{Figure_1}
\end{figure}

The surface in Fig.\ref{Figure_1} (left) shows the exponent of the covariance matrix eigenspectrum as a function of the parameters $(\rho, \varepsilon)$, for an ESN presented with the same set of natural images used in \cite{Stringer2019}. First of all, notice how the boundary exponent for ``smooth" high-dimensional input representations, $\mu \approx 1$, coincides with the transition between the stable an unstable regimes ($MLE = 0$, color-coded). This means that the system is able to generate ``optimal" representations of the input images, with properties akin to those observed in the visual cortex of mice, provided its dynamics lays close to a critical point!

Now, do these representations translate into a better performance when the system has to carry an input-processing task? To answer this question we relied on a classification task over the MNIST dataset, a common benchmarking dataset consisting in hand-written digits. In Fig.\ref{Figure_1} (right) we analyzed the percentage of images that were correctly classified during the test as a function of the spectral radius, while also recording the MLE as a measure of distance to the edge-of-instability. Remarkably, optimal performance for this classification task is obtained when the system is approaching the edge-of-instability from the stable regime, concomitant with a exponent $\mu \approx 1$ for the eigenspectrum of the activity covariance matrix.

To conclude, it is important to remark that the inputs fed to the network to obtain the previous results were natural images and MNIST digits, and thus can be considered high-dimensional (i.e., $d\rightarrow\infty$), so that the upper bound in the covariance spectrum exponent for the smoothness of the manifold is given by $\mu\geq 1+2/d\approx1$. Should we expect to find a similar match between edge-of-instability dynamics and optimal, smooth representations of low-dimensional inputs?

\begin{figure*}
\centering{}\includegraphics[width=\textwidth]{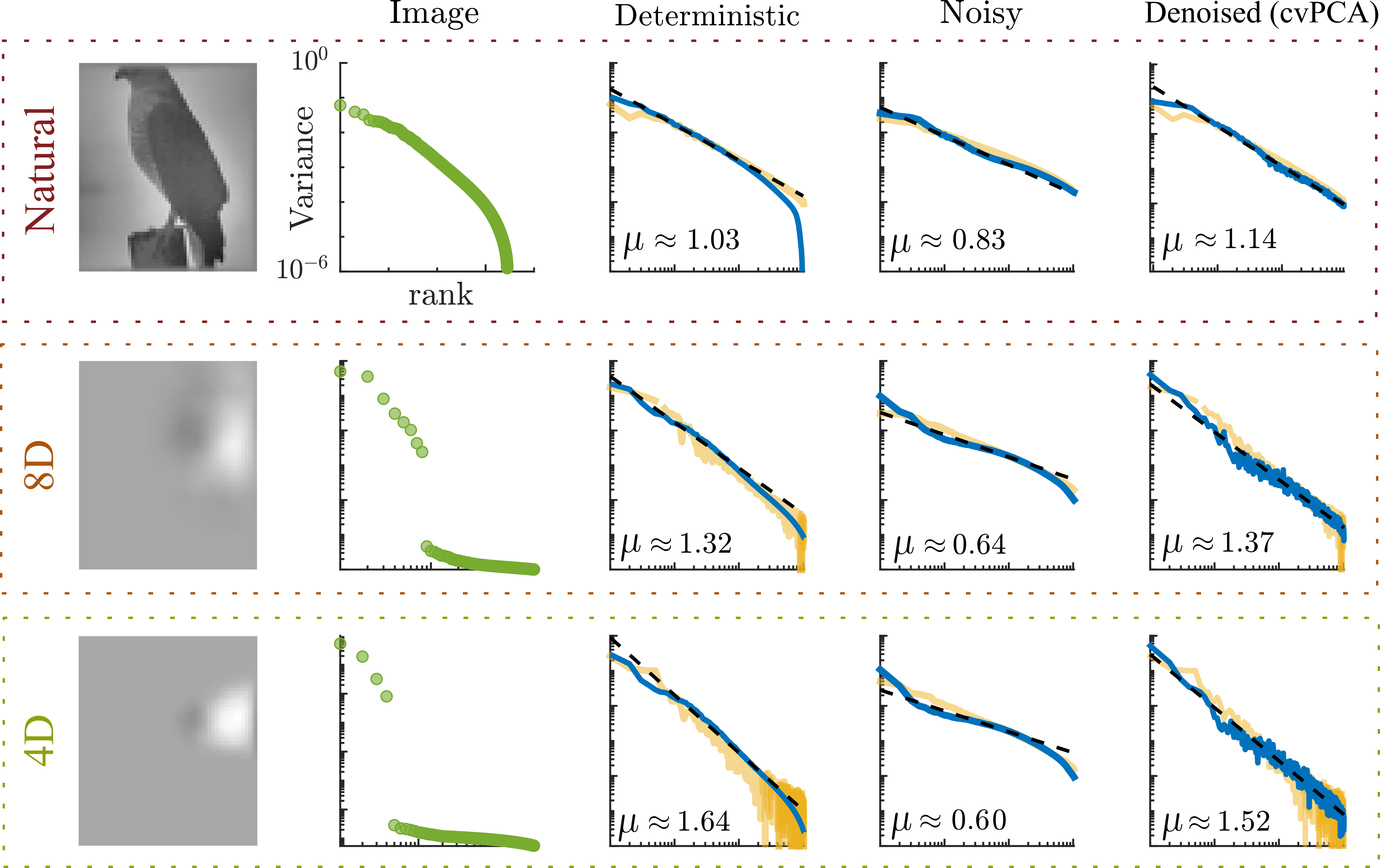}
\caption{\textbf{ESNs near the edge-of-instability generate representations with the same eigenvalue distributions observed in real neurons.} From left to right: one sample from the $M=2800$  images in the training set; covariance eigenspectrum of the images pixel intensities; covariance eigenspectrum of the internal states of an ESN (blue line) and real, V1 mouse neurons (yellow line, plotted after  \cite{Stringer2019} applying cvPCA) when subject to images of dimensionality $d$ ; same analysis, but now zero-mean white noise of amplitude $\epsilon = 0.4$  is added to the neuron dynamics (blue line), and no cvPCA is performed over the experimental recordings (yellow line); same analysis as in the previous panel, but now noise has been sustracted using cvPCA in both, simulations and experiments. From top to bottom: results for natural, high-dimensional images; the same images projected onto 8 dimensions; the same images projected onto 4 dimensions. To obtain the ESNs eigenspectra, parameters were chosen so that the networks operated near the edge-of-instability, with MLE $\sim -5 \cdot 10^{-3}$. Adapted with modifications from \cite{morales_optimal_2021}.}\label{Figure_2}
\end{figure*}

Following the experimental protocol of Stringer \emph{et al.}, we constructed $8$-dimensional and $4$-dimensional inputs by projecting the natural images used in \cite{Stringer2019} into their first $d=8$ and $d=4$ principal component, respectively. Fig.\ref{Figure_2} shows the rank-ordered covariance eigenvalues of the reservoir states, together with the best power-law fit exponents, for natural and low-dimensional inputs. For the sake of comparison, we also reproduced in yellow the eigenvalue rank-ordered distributions obtained from the experimental data on mouse visual cortex \cite{Stringer2019}. Not only our simulations are in perfect agreement with the experimental results, but one can also add a small noise $\mathbf{\xi}\neq0$ into the internal dynamics of the reservoir units, recovering the observed exponents before isolating the input-related variance through a cross-validated PCA (cvPCA) analysis, as explained in Section 5.1.

Therefore, summing up, the artificial neural network that we have designed under the paradigm of reservoir computing is able to construct optimal representation of the inputs when it operates close (but slightly below) to criticality (the edge of instability). It is  precisely under such conditions the network performance is optimal.

\section{Summary, discussions, and open directions}
\label{sec:final}

We have presented a large diversity of analyses of neural networks, both biological and artificial, trying to shed light on the relationships between their internal dynamical regimes and their emergent or collective computational capabilities.

First of all, we observed the scale-invariant nature of neural activity in the mouse brain emerge under a phenomenological renormalization group flow, resembling closely what happens at critical points of second order phase transitions. Moreover, the so-determined scaling exponents exhibit little (yet some) variability across brain regions, so we referred to them as been "quasi-universal", in seeming analogy with universality in critical phenomena.


After these scaling analyses, we set out to quantify the actual distance of empirical data to the edge of instability, finding that all 16 regions analyzed turn out to be relatively close to the edge of instability. Interpreting the reason behind the variability in the distance to criticality across regions remains an open and challenging goal. In particular, we have speculated that this variability could be related to a hierarchy of information flow in the brain organization 
\cite{morales_quasiuniversal_2023}.

\cite{morales_quasiuniversal_2023}.

To close the loop we constructed a very simple model of 
firing neurons, which can exhibit a phase transition as the overall coupling strength is increased. We found that, quite surprisingly, this simple model ---when tuned to the vicinity of its critical point--- exhibits scaling as elucidated by the phenomenological renormalization group technique. Moreover, the scaling exponents are surprisingly close to 
those found in actual brain activity. This also points out in the direction of possible universality. Analyzing whether this behavior is really universal in the strict sense of statistical physics or if it is approximate remains as an open challenge to be clarified in future work.

In summary, we report on solid quantitative evidence that actual neural activity in the mouse brain seems to operate, rather generically, close to a critical point or point of instability. Thus, the next question to be posed is: what is the reason for that?

Even if the potential functional advantages of critical behavior in  both real and artificial neural networks \cite{Plenz-functional, boedecker_information_2011,Hidalgo} (as well as, in general, on biological systems,  \cite{Mora,RMP}), have been extensively explored in recent years, here we go a step further and aimed at finding a relation between critical behavior and optimal representation of external inputs, i.e., optimal computational capabilities.

For this, we first recapitulated a recent seminal contribution by Stringer \emph{et al.} \cite{Stringer2019}. They first devised a method to separate variance stemming from inputs (signal covariance) from intrinsic fluctuations (noise covariance). Then, they derived a rather elegant mathematical theorem establishing that optimal representations of inputs within a network of neurons should be constructed on lower-dimensional manifolds in the space of neural activity, in which the corresponding spectrum of the (signal) covariance matrix should decay with an exponent ($\mu$) slightly larger that $1$, which is larger for lower-dimensional (simpler) inputs (i.e. $\mu \approx 1+2/d$, where $d$ is the input dimensionality).

All this might seem very abstract, however, it provides us with a quantitative tool to scrutinize how close to optimality are actual patterns of neural activity while encoding inputs. Indeed, by analyzing high-quality data from  the mouse visual cortex while the mouse was exposed to a series of visual stimuli, Stringer \emph{et al.} achieved the very remarkable conclusion that, as a matter of fact, empirical data describe neural representations that are very close to the predicted optimal behavior, i.e., the spectrum of the signal covariance decays with an exponent slightly larger than one, that increases as the set of stimuli becomes simpler (lower dimensional).

Therefore, on the one hand, we have found evidence of scaling and close to critical behavior across brain areas and, on the other hand, we have found that the visual cortex seems to operate in a regime allowing it to generate optimal representations of stimuli.
Thus, we wondered whether these two features come 
generically hand by hand.

To shed light onto this question we constructed an artificial neural network; more precisely an 
echo-state network, which is one of the best known examples of reservoir computing. This is a paradigm of computation different from the standard von-Neumann's one, allowing to exploit the collective dynamical state of the neural network (the reservoir) in a generic 
and flexible way, without modifying synaptic weights but by training readouts of the overall network state. 

The internal state of the network can be tuned by changing some global parameter (such as e..g the overall synaptic strength) and there is a critical point where the dynamics becomes unstable. We found that
right at such a critical point, the network exposed to
series of inputs (actually, the very same images used by Stringer \emph{et al.} in \cite{Stringer2019}), the network
produces optimal representations (i.e. $\mu \approx1$) when it is tuned to the edge of instability. 
Moreover, and rather remarkably, the performance of the network when trained to perform a specific task
(classification of digits from the MNIST datasets) is found to be optimal when the network is slightly below the edge of instability, i.e. the regime where the optimal condition for the covariance spectrum is verified.

The corollary of all this is that neural networks ---both biological and artificial--- seem to behave in an optimal fashion in what respects input representation and computational performance if they are tuned to be slightly below the edge of instability, i.e. when they are close to criticality.  This result adds to the  mounting evidence revealing that criticality is a key concept for the understanding of neural networks as well as other 
information-processing biological systems \cite{RMP}.

As a matter of fact, in recent years, there have been some attempts  to design general-purpose neuromorphic computation devices (chips) exploiting the potential advantages of criticality (see, e.g., \cite{Stieg,Kuncic,Cramer2020,Brown}), a direction that we find most suggestive and promising.

We hope that more synergies between 
neuroscience and machine learning
emerge from these analyses. These should help us deciphering the meaning and functionality of brain activity as well as designing novel and powerful artificial-intelligence paradigms inspired more closely by biological networks.

\section*{Appendix A: Relating covariances to connectivity}
\addcontentsline{toc}{section}{Appendix}

To prove Eq.(\ref{CofJ}) for the linear-rate model, one just needs to take the trial-averaged (average over realizations of $\xi_i(t)$) firing rate at the stationary state, in the frequency domain:
\begin{equation}
\langle x_i(\omega)\rangle =(I-J)^{-1}\langle \xi_i(\omega)\rangle,
\end{equation}
where $\langle \cdot \rangle$ denotes an average over trials. Then the matrix $C(\omega)$, i.e. the matrix of the Fourier transform of the covariances, is given by:
\begin{equation}
C_{ij}(\omega)=\langle x_i(\omega)x_j(\omega)\rangle = (I-J)^{-1}((I-J)^{-1})^T
\end{equation}

This relation ---that has been proven to be remarkably general--- holds as a very good approximation for many different (nonlinear) models, including spiking neural networks (see \cite{Pernice,Trousdale,Dahmen2019} and \cite{Shea-Brown} for a recent review). For our purposes, we limited the analysis to the case in which $\omega=0$, for which $C_{ij}$ can be easily interpreted as the long-time window or spike-count covariance.

\section*{Appendix B: Rank-ordered eigenvalues vs probability densities}
\addcontentsline{toc}{section}{Appendix}

In Section \ref{sec:RG} we argued that when correlations are self-similar inside a cluster of size $K$, the $n$-th largest eigenvalue $\lambda_{n}$ decays with its rank $n$ (when ordered from  the largest to the smallest) as:
\begin{equation}
\lambda_{n} = A \left(\frac{n}{K}\right)^{-\mu}.
\label{eq:rankord}
\end{equation} 

On the other hand, in Section \ref{yuhu} we reported on the probability density of the eigenvalues of the covariance matrix (Eq.\ref{eq:pRC}), and argued that its tail converges to a power-law 
when the system is close to the edge of instability: 
\begin{equation}
p(\lambda) \sim B \lambda^{-\nu}.
\label{eq:pRCc}
\end{equation}
Here we make explicit the relationship between these two power-laws.

The spectrum of a matrix of size $K$ is composed by $K$ discrete eigenvalues:
\begin{equation}
    p(\lambda)=\frac{1}{K}\sum_{n=1}^K \delta\left(\lambda-\lambda_{n}\right).
\end{equation}
In the limit of $K\rightarrow\infty$ the above discrete distribution can be approximated by a continuous one:
\[
p^{(c)}_{RC}(\lambda)\approx\frac{1}{K}\int_{1}^{\infty}\delta\left(\lambda-\lambda_{n}\right)dn = \frac{1}{K}\int_{1}^{\infty}\delta\left(\lambda-A \left(\frac{n}{K}\right)^{-\mu}\right) dn,
\]
where we used Eq.\ref{eq:rankord} for the rank-ordering of the spectrum . The function $g(n)\equiv \lambda-A\left(\frac{n}{K}\right)^{-\mu}$ has roots $n_{0}=K\left(\frac{\lambda}{A}\right)^{-\frac{1}{\mu}}$. Using the following property of the Dirac-delta distribution:
\begin{equation}\delta\left(g(n)\right)=\dfrac{\delta\left(n-n_{0}\right)}{\left|g'\left(n_{0}\right)\right|}=\dfrac{\delta\left(n-n_{0}\right)}{K^{-1}A\mu \left(\frac{\lambda}{A}\right)^{\frac{\mu+1}{\mu}}},
\end{equation}
the spectral density reads:
\begin{equation}
p^{(c)}_{RC}(\lambda)\approx\frac{1}{K \mu} A^{\frac{1}{\mu}}K\lambda^{-1-\frac{1}{\mu}} \int_{1}^{\infty}\delta\left(n-n_{0}\right)dn=\frac{1}{\mu} A^{\frac{1}{\mu}}\lambda^{-1-\frac{1}{\mu}}
\label{eq:pRC2}
\end{equation}

Thus, by comparing Eq. \ref{eq:pRC} with Eq.\ref{eq:pRC2}, one readily finds a relationship between the power law exponent $\mu$ of the rank-ordered eigenvalues and the exponent of the spectral distribution $\nu$:
\begin{equation}
\nu=1+\frac{1}{\mu}.
\label{expomunu}
\end{equation}
The reader can find a more detailed discussion of this scaling relation in,  e.g.,\cite{Li2002,DeMarzo2021}.


\section*{Appendix C: Cross-validate PCA}
\addcontentsline{toc}{section}{Appendix}

Following Stringer {\emph{et al.}} \cite{Stringer2019}, let us consider two observation matrices, $X_{(1)}, X_{(2)}\in\mathbb{R}^{N\times T}$, corresponding to two identical realizations or trials of an experiment. For simplicity, we further assume that the mean activity of each neuron across images has been subtracted, so that both matrices have zero-mean rows. The Singular Value Decomposition (SVD) theorem states that any observation matrix $X$ can be decomposed as:
\begin{equation}
X  =USV^{T},
\end{equation}
so that 
\begin{equation}
XV  =US\Longrightarrow X\mathbf{v}_{i}=\sigma_{i}\mathbf{u}_{i},
\end{equation}
where $U\in\mathbb{R}^{N\times r}$ contains the $\mathbf{u}_{i}$
eigenvectors by columns, $V\in\mathbb{R}^{T\times r}$ contains the
$\mathbf{v}_{i}$ eigenvectors by columns, and $S\in\mathbb{R}^{r\times r}$
is a diagonal matrix containing the singular values $s_{ii}=\sigma_{i}=\sqrt{\lambda_{i}}$.

Assume one has performed SVD on the first trial observation matrix $X_{(1)}$, obtaining the singular vectors $U_{(1)}$ and $V_{(1)}$. 
Defining the projection matrix $P=U_{(1)}^{T}\in\mathbb{R}^{r\times N}$, which diagonalizes $C_{u}^{(1)}=\frac{1}{T-1}X_{(1)}X_{(1)}^{T}$, one can compute:

\begin{equation}
Y_{(1)}=PX_{(1)}\Longrightarrow\mathbf{y}_{i}^{(1)}=(\mathbf{u}_{i}^{(1)})^{T}X_{(1)}\in\mathbb{R}^{1\times T},
\end{equation}
where $Y_{(1)}\in\mathbb{R}^{r\times T}$ contains the projection of the first trial activity over its $r$ principal components. Now, the idea of cv-PCA is to project on the same subspace the observations coming from the second trial \cite{Stringer2019}:
\begin{equation}
Y_{(2)}=PX_{(2)}\Longrightarrow\mathbf{y}_{i}^{(2)}=(\mathbf{u}_{i}^{(1)})^{T}X_{(2)}\in\mathbb{R}^{1\times T},
\end{equation}
and ask to form a covariance matrix with the product of both projections:
\begin{equation}
C_{\psi}=\dfrac{1}{T-1}Y_{(1)}Y_{(2)}^{T}=\dfrac{1}{T-1}\left(\begin{array}{ccc}
\mathbf{y}_{1}^{(1)}\cdot\mathbf{y}_{1}^{(2)} & \mathbf{y}_{1}^{(1)}\cdot\mathbf{y}_{2}^{(2)} & ...\\
\mathbf{y}_{2}^{(1)}\cdot\mathbf{y}_{1}^{(2)} & \mathbf{y}_{2}^{(1)}\cdot\mathbf{y}_{2}^{(2)} & ...\\
... & ... & \mathbf{y}_{r}^{(1)}\cdot\mathbf{y}_{r}^{(2)}
\end{array}\right).
\end{equation}
One can now hypothesize that the activity at time $t$ of any neuron $i$ during trial $k$ can be linearly decomposed as:
\begin{equation}
\mathbf{x}_{i}^{(k)}(t)=\mathbf{\psi}_{i}(t)+\mathbf{\epsilon}_{i}^{(k)}(t,\label{Decomposition}
\end{equation}
where $\mathbf{\psi}_{i}$ denotes the input-related activity, which should be independent of the trial for experiments carried in identical conditions; and $\mathbf{\epsilon}_{i}^{(k)}$ is defined as background or trial-to-trial variable activity, spanning a subspace that is orthogonal to the input-related one. Let us now show that the i-th diagonal element of $C_{\psi}$ is a non-biased estimator of the input-related variance $\omega_{i}$:
\begin{equation}
\omega_{i}=\dfrac{1}{T-1}\mathbf{y}_{i}^{(1)}\cdot\mathbf{y}_{i}^{(2)}=\dfrac{1}{T-1}(\mathbf{u}_{i}^{(1)})^{T}X_{(1)}X_{(2)}^{T}\mathbf{u}_{i}^{(1)}.
\end{equation}
Rewriting the observation matrices as $X_{(k)}=\Psi+\Sigma_{(k)}$,
where $\Psi$ contains vectors $\mathbf{\psi}_{i}$ in rows and
so does $\Sigma_{(k)}$ with vectors $\mathbf{\epsilon}_{i}^{(k)}$:
\begin{equation}
\omega_{i} =\dfrac{1}{T-1}\mathbf{u}_{i}^{T}\left(\Psi+\Sigma_{(1)}\right)\left(\Psi+\Sigma_{(2)}\right)^{T}\mathbf{u}_{i} =\dfrac{1}{T-1}\mathbf{u}_{i}^{T}\Psi\Psi^{T}\mathbf{u}_{i}+\dfrac{1}{T-1}\mathbf{u}_{i}^{T}\Sigma_{(1)}\Psi^{T}\mathbf{u}_{i},
\end{equation} 
where all the terms containing $\Sigma_{(2)}$ can be dismissed due to the statistical independence of the realizations. Under reasonable approximations shown in \cite{Stringer2019}, one can prove that if the singular vectors $\mathbf{u}_{i}$ approach the singular vectors of $\Psi$, then the first term converges to the actual variance of the input-related activity along the principal direction $\mathbf{u}_{i}$, while the second term converges to zero. We have thus seen how the
\emph{cross-validated PCA} (cvPCA) method proposed by Stringer \emph{et al.} allows one to estimate the input-related covariances.

Is it possible to take a step further and find a proxy for the input-related activity observation matrix $\Psi\in\mathbb{R}^{N\times T}$, such that $\tilde{C}_{\psi}=\frac{1}{T-1}\Psi\Psi^{T}$ has the same eigenvalue spectrum as $C_{\psi}$? To do so, we first define a new basis of projected vectors given by:
\begin{equation}
\mathbf{z}_{i}=\sqrt{\mathbf{y}_{i}^{(1)}\cdot\mathbf{y}_{i}^{(2)}}\dfrac{\mathbf{y}_{i}^{(2)}}{\left\Vert \mathbf{y}_{i}^{(2)}\right\Vert }\in\mathbb{R}^{1\times T}.
\end{equation}
which clearly fulfills $\mathbb{E}\left[\mathbf{z}_{i}\cdot\mathbf{z}_{i}^{T}\right]=\omega_{i}$
under the same conditions as above. If $Z\in\mathbb{R}^{r\times T}$ is the matrix composed by the $r$ vectors $\mathbf{z}_{i}$ in rows, then:
\begin{equation}
\Psi=P^{T}Z.
\end{equation}
is the input-related activity of the neurons. From there, it is straightforward using  Eq.\ref{Decomposition} that one can estimate the background activity just by subtracting the input-related activity from the raw data:
\begin{equation}
\Sigma_{(k)}=X_{(k)}-\Psi.
\end{equation}

\vspace{1cm}
{\textbf{Acknowledgments:}}
We acknowledge the Spanish Ministry and Agencia Estatal de investigaci{\'o}n (AEI) through Project of I+D+i Ref. PID2020-113681GB-I00, financed by 
MICIN/AEI/10.13039/501100011033 and FEDER "A way to make Europe", as well as the Consejer{\'\i}a de Conocimiento, Investigaci{\'o}n Universidad, Junta de Andaluc{\'\i}a and European Regional Development Fund, Project reference P20-00173 for financial support. We also thank R. Calvo, C. Martorell, V. Buend{\'\i}a, P. Villegas, R. Corral, J. Pretel, P. Moretti, and M. Iba{\~n}ez, for  valuable discussions and/or suggestions.

\bibliography{biblio-morales.bib}

\end{document}